\documentclass[american,aps,pra,reprint]{revtex4-1}
\usepackage{newlfont}
\usepackage{color}           % color to the text 
\usepackage{soul}            % strickthrough the text
\usepackage{amssymb}
\usepackage{amsfonts}
\usepackage{amsmath}
\usepackage{wasysym}
\usepackage{graphicx}
\usepackage{epsfig}
\usepackage{amsthm}
\usepackage{bm}
\usepackage[colorlinks=true, citecolor=blue, urlcolor=blue ]{hyperref}
\usepackage{times}

\newcommand{\etal}{\textit{et al.}}

\begin{document}

%\title{Freezing of quantum discord in dynamic evolution of non-dephasing tunable quantum spin chains}
\title{Adiabatic freezing of long-range quantum correlations in spin chains}

\author{Himadri Shekhar Dhar, Debraj Rakshit, Aditi Sen(De), and Ujjwal Sen}
\affiliation{Harish-Chandra Research Institute, Chhatnag Road, Jhunsi, Allahabad 211 019, India}

%%\title{Adiabatic freezing of long-range quantum correlations in spin chains}
%%%\shorttitle{Title} %Insert here a short version of the title if it exceeds 70 characters
%%
%%\author{Himadri Shekhar Dhar, Debraj Rakshit, Aditi Sen(De), \and Ujjwal Sen}
%%\shortauthor{H. S. Dhar \etal}
%%
%%
%%\institute{Harish-Chandra Research Institute, Chhatnag Road, Jhunsi, Allahabad 211 019, India}
%%\pacs{03.67.-a} {Quantum information}
%%\pacs{03.67.Bg} {Entanglement production and manipulation}
%%\pacs{75.10.Jm} {Quantized spin models, including quantum spin frustration}

\begin{abstract}%Establishing long-range quantum correlation is an integral part of several quantum information processing tasks. 
We consider a process to create
quasi long-range quantum discord between the non-interacting end spins of a quantum spin chain, with the end spins weakly coupled to the bulk of the chain. 
%We find that 
The process is not only capable of creating long-range quantum correlation but
the latter remains frozen, when certain weak end-couplings are adiabatically varied
%, provided they are 
below certain thresholds.
We term this phenomenon as adiabatic freezing of quantum correlation.
We observe that the freezing is robust to moderate thermal fluctuations and {is intrinsically related to the cooperative properties of the quantum spin chain. In particular, we}
find that the energy gap of the system remains frozen for these adiabatic variations, 
{and moreover,} 
considering the end spins as probes, we show that the {interval of freezing} can detect the anisotropy transition in quantum \emph{XY} spin chains. {Importantly, the adiabatic freezing of long-range quantum correlations can be simulated with contemporary experimental techniques.}
\end{abstract}

\maketitle
\section{Introduction}
Quantum correlation \cite{ent,qcor} is one of the principal characteristics that separates the quantum domain 
from its classical counterpart. The properties and phenomena that arise from it are neither reproducible nor simulatable in  classical systems.
The amount of quantum correlation that exists between two subsystems of a pure quantum state is completely captured by entanglement \cite{ent}. 
However, for mixed states, local measurements may reveal nonclassical features, that are present even in non-entangled or separable states \cite{qcor}. 
Measures that capture quantum correlations beyond entanglement, such as quantum discord (QD) \cite{disc}, have been instrumental in investigating several protocols of 
quantum information and computation (QIC) \cite{comp,comp1,comp2}, quantum phase transitions \cite{qpt}, many-body dynamics \cite{mb,epl}, quantum biology \cite{qbio}, and metrology \cite{qmet,qmet1} (see \cite{qcor}, for a review).

Methods to create long-range entanglement have  attracted a lot of attention due to their importance in several protocols in QIC \cite{swap,repeat,local,grudka,venuti,ilu1,bose,referee}.
Such investigations led to the discovery of processes like entanglement swapping \cite{swap} and repeaters \cite{repeat}, 
%entanglement concentration \cite{concentrate}, 
and concepts like localizable entanglement \cite{local}. In the last decade, there have been several instances
where quantum correlation measures such as QD were claimed to be important \cite{qcor,comp,comp1,comp2,qpt,mb,epl,qbio,qmet,qmet1}. 
%
% Immense importance of quantum discord also demands procedures to prepare long-range quantum discord in realizable systems.
%
For example, it 
provides an interesting perspective on the dynamics of open quantum systems  \cite{open1}, where entanglement is fragile. In quantum systems subject to noisy environments, 
for both Markovian and non-Markovian evolution, QD is more robust than entanglement \cite{robust,esd}. %Further, for certain disspative processes, entanglement is known to vanish in finite evolution times, a phenomena known as entanglement sudden death (ESD) \cite{esd}, while QD remains finite. 
Interestingly, for certain types of quantum states under dephasing, QD exhibits a qualitatively different robustness. It remains \emph{frozen} for finite evolution times \cite{manis}, even though entanglement suffers sudden death. In recent times, a significant amount of research has been undertaken to characterize the {phenomena of 
freezing of quantum correlations \cite{expt,freezing, titas}, including entanglement \cite{freeze-ent}}.

%The increasing importance of QD demands the investigation of procedures to prepare long-range QD in realizable systems.
In this letter, {we report %freezing of 
quasi long-range QD} between the end spins of a finite quantum spin chain, when the end spins are weakly coupled to the bulk of the spins. 
{The work is motivated by the desire to investigate the quantum correlation properties of the end spins due to the collective effect of the bulk spins that act as a reservoir for the end spins, thus giving rise to interesting long-range quantum phenomena.}
%See Fig. \ref{fig:0}.
%It is known that under weak end-coupling, the end spins can be entangled \cite{grudka,venuti,zan,new}, giving rise to quasi long-range entanglement in a spin model with short-range interaction. 
%The long-range entanglement is maximal for extremely  weak end-coupling and decreases with increase in the value of the latter and also the system size, exhibiting a non-temporal ESD. 
%However, we observe, that the long-range quantum discord between the end spins is more robust, with respect to variation in system parameters, as compared to long-range entanglement. 
%%Interestingly, if the weak end-coupling is varied along the different spin components (say, $X$-- and $Y$--axis) of the end spins, the long-range quantum discord is frozen under the increase of one the weak coupling components, with a freezing length that is commensurate with the other component. 
%
In particular, we find that the long-range QD of the ground state can exhibit non-temporal freezing, while the weak end-couplings are varied. 
We term the {freezing
% of long-range QD 
as} ``adiabatic'', as the phenomenon is observed while the weak end-couplings 
are adiabatically varied.
%%%Interestingly, the freezing interval, i.e.,
%%%the region of weak end-coupling for a certain spin-component over which the quantum correlation is frozen, is commensurate with the value of the end-coupling along another spin-component.
%
No such freezing behavior is observed for the long-range entanglement. 
{The observed phenomena 
%of adiabatic freezing 
makes long-range QD a robust resource, against moderate thermal fluctuations and variations in certain system parameters, for implementing 
%quantum mechanical 
quantum protocols between distant spin qubits.}

{Interestingly, and in contradistinction to temporal freezing \cite{manis, expt,freezing, titas},
% \cite{manis,expt,freezing, sarandy, titas}, 
the observed %freezing 
adiabatic phenomenon is characteristic of the quantum system rather than an external environment, and is intrinsically related to the many-body properties of the quantum spin chain.}
%%%Moreover, we find that there exists a complementary relation between the frozen discord value and the weak coupling. 
%can be increased by suitably tuning system parameter and size with the loss of freezing length.
%
%
%%\begin{figure}[t]
%%%\begin{center}
%%\epsfig{figure= spincoupling.eps,
%%%try3.eps, 
%%height=.13\textheight, width=0.42\textwidth,angle=0}
%%%\epsfig{figure= try2.eps}
%%\caption{(Color online.) The set-up. The end spins are weakly coupled (black-dashed lines) with the bulk,
%%%in the quantum spin chain with $N$ spins. 
%%%The ends spins (red circle) are weakly coupled (black-dashed lines) to the bulk. 
%%which are strongly coupled to each other (black solid line). The frozen discord is observed between spins $1$ and $N$. 
%%} 
%%\label{fig:0}
%%%\end{center}
%%\end{figure}
%
%
%distantly separated spins of a chain. 
%
%%{This is evident from the ability of the adiabatic freezing of QD to lend itself for investigation of certain important properties of quantum spin chains.}
%%
%%
%%
%%{An}
%%%%Further 
%%important feature of the adiabatic freezing of QD lies in its ability to 
%%%serve as a figure of merit for 
%%lend itself for investigation of certain important properties of quantum spin chains. 
{For instance, in an experimental setting, one may consider the two end-spins to be 
%controlled 
\emph{probe} sites weakly coupled to a \emph{system} consisting of a spin chain \cite{venuti,ilu1, bose}, where the weak end-couplings can be controlled. }
The long-range QD between the probe spins can then be shown to identify the nature of interaction in the system. Specifically, the freezing interval of long-range QD can be used to define an order parameter that detects the ``anisotropy transition'' in the system considered.
{Further, one can show that the adiabatic freezing phenomena is not limited to long-range quantum correlations but is also manifested in other cooperative properties of the quantum spin chain, such as the energy gap.}
%
%We also observe that the energy gap in the considered system undergoes adiabatic freezing.
% 

{We note that there are recent experimental proposals and techniques to generate and characterize long-range quantum correlation in quantum spin systems \cite{new,iontrap,flux} (also see Refs.~\cite{expt3}).
In this light, we investigate the effect of thermal fluctuations in the spin system, which is important for potential experimental implementations, and find that the adiabatic freezing is stable below a critical temperature. Hence, our results show that in an experimentally accessible regime of weak end-couplings and low-temperature, high quasi-long range quantum correlations can be adiabatically frozen for applications in various quantum information protocols.} 

\section{Methodology}  
\label{method}
Let us consider an anistropic \textit{XY} quantum spin chain containing $N$ spins with a closed end. The Hamiltonian for such a system can be written as
\begin{equation}
\cal{H} = \sum_i^N \frac{\kappa}{4}~(\cal{J}_i~ \sigma^x_i \sigma^x_{i+1} + \cal{K}_i ~\sigma^y_i \sigma^y_{i+1}), 
\label{Ham}
\end{equation}
where $\sigma^{x(y)}_{N+1}$ = $\sigma^{x(y)}_{1}$.
%, satisfies the desired boundary condition. 
%
$\cal{J}_i$ and $\cal{K}_i$ are the dimensionless interaction strengths.
$\kappa$($> 0$) has the unit of energy. 
% between the nearest-neighbors $i$ and $i+1$, along the $X$-- ($Y$--) direction. 
$\sigma^{i}$, $i = x, y, z$, are the Pauli spin matrices.
The open-end case is obtained by setting $\cal{J}_N$ and $\cal{K}_N$ equal to zero. 

The quantum correlation between two arbitrary sites, in the ground and thermal equilibrium state of the Hamiltonian can be obtained by deriving the two-site reduced density matrix, following the seminal work in \cite{LSM}. 
{The symmetry of the Hamiltonian ensures that all (single-site) magnetizations $\langle \sigma^\alpha_i\rangle$, $\forall~\alpha =(x,y,z)$ vanish in the absence of any
%. However, for no 
external fields.
%, the single-site magnetization also vanishes. 
The only non-vanishing two-site terms are $\langle \sigma^x_i \sigma^x_j \rangle$, $\langle \sigma^y_i \sigma^y_j \rangle$, and $\langle \sigma^z_i \sigma^z_j \rangle$. %, which gives us the two-site correlation functions.
%, since $\langle \sigma^\alpha_i\rangle$ = 0, $\forall~\alpha =(x,y,z)$}.
Hence, any two-site reduced density matrix, for arbitrary sites $i$ and $j$, can be written as
%\begin{equation}
$\rho_{ij}$ = $1/4(\mathbb{I} + \sum_{\alpha = x,y,z} T^{\alpha\alpha}_{ij}~ \sigma^\alpha_i \otimes \sigma^\alpha_j)$,
%\label{rho12}
%\end{equation}
where $T^{\alpha\alpha}_{ij}$ = $\langle \sigma^\alpha_i \sigma^\alpha_j \rangle$ are the two-site correlation functions and $\mathbb{I}$ is the two-qubit identity matrix. 
%
%In terms of the correlation matrix, $\mathcal{G}$, 
$T^{\alpha\alpha}_{ij}$ can be analytically derived by solving the Hamiltonian in Eq.~(\ref{Ham}) \cite{LSM,supple}.
Since $\rho_{ij}$ is Bell-diagonal, its QD \cite{disc} can be analytically obtained \cite{LUO}. 
%For the Bell-diagonal density matrix, $\rho_{ij}$, its eigenvalues, $e_i$, can be obtained in terms of the two-site correlation functions as
The eigenvalues of $\rho_{ij}$, in terms of $T^{\alpha\alpha}_{ij}$, is given by $\{e_i\}$ equal to $1/4(1 \pm (T^{xx}_{ij} + T^{yy}_{ij}) - T^{zz}_{ij})$ and $1/4(1 \pm (T^{xx}_{ij} - T^{yy}_{ij}) + T^{zz}_{ij})$, for $i$ = 1 to 4.
%%\begin{eqnarray}
%%e_1 &=& 1/4 - T^{xx}_{ij} - T^{yy}_{ij} - T^{zz}_{ij};\nonumber\\
%%e_2 &=& 1/4 - T^{xx}_{ij} + T^{yy}_{ij} + T^{zz}_{ij};\nonumber\\
%%e_3 &=& 1/4 + T^{xx}_{ij} - T^{yy}_{ij} + T^{zz}_{ij};\nonumber\\
%%e_4 &=& 1/4 + T^{xx}_{ij} + T^{yy}_{ij} - T^{zz}_{ij}.
%%\end{eqnarray}
%$
%e_1$ = $1/4 - T^{xx}_{ij} - T^{yy}_{ij} - T^{zz}_{ij}$;
%$e_2$ = $1/4 - T^{xx}_{ij} + T^{yy}_{ij} + T^{zz}_{ij}$;
%$e_3$ = $1/4 + T^{xx}_{ij} - T^{yy}_{ij} + T^{zz}_{ij}$;
%$e_4$ = $1/4 + T^{xx}_{ij} + T^{yy}_{ij} - T^{zz}_{ij}$.
%
The quantum mutual information is given by the relation, $\cal{I}(\rho_{ij})$ = $\sum_i e_i \log_2 (4e_i)$, and the classical correlation, obtained after optimization over measurements on a single-party, is given by 
$\cal{C}(\rho_{ij})$ = $\sum_{k=1}^2 x_k \log_2 (2x_k)$, where $x_k$ = $(1$ + $(-1)^k ~x)/2$ ($k$ = 1,2), and 
$x$ = $\max\{\lvert T^{xx}_{ij}\rvert,\lvert T^{yy}_{ij}\rvert,\lvert T^{zz}_{ij}\rvert\}$.
%%\begin{eqnarray}
%%\cal{C}(\rho_{ij}) &=& \sum_{k=1}^2 x_k \log_2 (2x_k), ~~\mathrm{where}\\
%%x_k &=& (1 + (-1)^k ~x)/2~~~~ (k = 1,2), \mathrm{and}\nonumber\\
%%x &=&  4 \max\{\lvert T^{xx}_{ij}\rvert,\lvert T^{yy}_{ij}\rvert,\lvert T^{zz}_{ij}\rvert\}.\nonumber
%%\end{eqnarray}
The QD is then given by the relation, 
%%\begin{eqnarray}
%%\mathcal{D}(\rho_{ij}) &=& \cal{I}(\rho_{ij}) - \cal{C}(\rho_{ij})\nonumber\\
%%&=& \sum_{i=1}^4 e_i \log_2 (4e_i) - \sum_{k=1}^2 x_k \log_2 (2x_k).
%%\end{eqnarray}
$
\mathcal{D}(\rho_{ij})$ = $\cal{I}(\rho_{ij}) - \cal{C}(\rho_{ij})
$ = $\sum_{i=1}^4 e_i \log_2 (4e_i) - \sum_{k=1}^2 x_k \log_2 (2x_k)$.
Similarly, the entanglement, using concurrence \cite{woot}, between any two sites can be analytically derived in terms of the correlation function $T^{\alpha\alpha}_{ij}$ \cite{supple}:
 %concurrence \cite{woot}. For the 
%Bell-diagonal density matrix, $\rho_{ij}$, the concurrence can be obtained using 
%the two-site correlation functions and is given by 
$
\cal{E}(\rho_{ij})$ = $\max\left[0, 2\max[\{e_i\}]-1\right].
$}
%where $g^{\pm}_{ij} = T^{xx}_{ij} \pm T^{yy}_{ij}$, and $h^{\pm}_{ij}$ = $1 \pm T^{zz}_{ij}$.

%%%%From the perspective of adiabatic freezing of quantum correlations, one can consider an $N$-spin open quantum spin chain, with nearest neighbor interactions, where the two spins at the edge of the chain (end spins) are weakly coupled to the remaining bulk of $N-2$ spins.
%%%%%, as shown in Fig.~\ref{fig:0}.
%%%%% through nearest neighbor interactions. 
%%%%The Hamiltonian for such a spin chain, can be written as the sum of a Hamiltonian, $\cal{H}_{bulk}$, containing the interaction of the central block of spins, and another, $\cal{H}_{end}$, containing the end spin interactions, and is given by, where%\label{eq:ham1}\\ 
%%%%\begin{eqnarray}
%%%%\cal{H} &=& \cal{H}_{bulk} + \cal{H}_{end} \\
%%%%\cal{H}_{bulk} &=& \sum_{i=2}^{N-2} \frac{\kappa}{4}~ (\cal{J}_i ~\sigma^x_i \sigma^x_{i+1} + \cal{K}_i~\sigma^y_i \sigma^y_{i+1}),\label{eq:ham2}\\ 
%%%%\cal{H}_{end} &=& \frac{\kappa}{4} \left[\lambda_1 (\sigma^x_{1} \sigma^x_{2} + \sigma^x_{N-1} \sigma^x_{N})\right.\nonumber\\
%%%%&+& \left.\lambda_2 (\sigma^y_{1} \sigma^y_{2} + \sigma^y_{N-1} \sigma^y_{N})\right].
%%%%\label{eq:ham3}
%%%%\end{eqnarray}
%%%%%where $\lambda_1 = \lambda_2 \le \left\{\cal{J}_i\right\}$, for balanced weak coupling. 
%%%%The above Hamiltonian can be exactly solved to obtain any two-site density matrix, using Eqs.~(\ref{rho12}) and (\ref{corr}), and thus calculate the two-spin quantum correlation.

\section{Adiabatic freezing of quantum discord}
Let us consider an $N$-spin non-periodic quantum spin chain, with nearest neighbor interactions, where the two spins at the edge of the chain (end spins) are weakly coupled to the remaining bulk of $N-2$ spins.
% through nearest neighbor interactions. 
The Hamiltonian for the spin is given by
% the sum of a Hamiltonian, $\cal{H}_{bulk}$, containing the interaction of the central block of spins, and another, $\cal{H}_{end}$, containing the end spin interactions, and is given by, 
$\cal{H} = ({\kappa}/{4})~(\cal{H}_{bulk} + \cal{H}_{end})$, where%\label{eq:ham1}\\ 
\begin{eqnarray}
&\cal{H}_{bulk} = \sum_{i=2}^{N-2} (\cal{J}_i ~\sigma^x_i \sigma^x_{i+1} + \cal{K}_i~\sigma^y_i \sigma^y_{i+1}),&\label{eq:ham2}\\ 
&\cal{H}_{end} = \lambda_1 (\sigma^x_{1} \sigma^x_{2} + \sigma^x_{N-1} \sigma^x_{N})+\lambda_2 (\sigma^y_{1} \sigma^y_{2} + \sigma^y_{N-1} \sigma^y_{N}).&
\label{eq:ham3}
\end{eqnarray}
%where $\lambda_1 = \lambda_2 \le \left\{\cal{J}_i\right\}$, for balanced weak coupling. 
The above Hamiltonian can be exactly solved to calculate the quantum correlation between any two spins in the chain.
% \cite{supple}.
%
The cases for which $\lambda_1 = \lambda_2 \le \cal{J}_i,\cal{K}_i$, $\forall i=2,...,N-2$, will be referred to as ``balanced'' weak-coupling.
{It is known that for $\lambda_1(=\lambda_2) \to 0$, %and when the entire chain is in the ground state, 
the two end spins in the ground state of the chain are maximally entangled}. Denoting concurrence \cite{woot} by $\cal{E}_L$, we have $\cal{E}_L \to 1$ in that limit. This gives rise to quasi long-range entanglement, interestingly, in a quantum system with very short-range interaction \cite{grudka,venuti}. In the balanced case, $\cal{E}_L$ decreases monotonically with increasing $\lambda_1$(= $\lambda_2$), and the {rate of decay increases with system size}. Also, 
%As $\lambda_1$ increases the sizable long-range quantum correlation monotonically decreases. 
$\cal{E}_L$ may exhibit non-temporal sudden death (cf. \cite{epl}).% and decreases at a faster rate for spin chains with higher number of spins \cite{zan}.
%%%
\begin{figure}[h]
\begin{center}
\includegraphics[width=6cm]{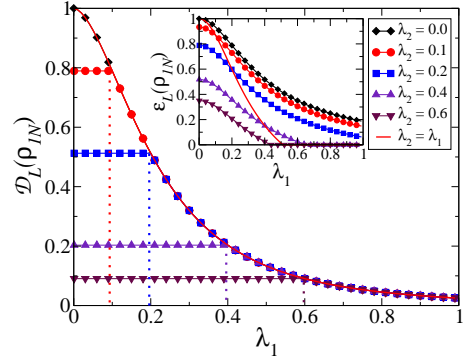}
\caption{(Color online.) Adiabatic freezing of quasi long-range quantum discord ($\cal{D}_L$). 
%%%The figure shows the %adiabatic dynamics of $\cal{D}_L$ for the spin chain, of $N = 10$ spins, with increasing end-spin coupling, 
%strength along the $X$--axis, 
%%%$\lambda_1$, for fixed values of the weak coupling,
% along the $Y$--axis, $\lambda_2$. 
{The bulk is an \emph{XX} spin chain}.
For each non-zero value of $\lambda_2$, freezing of $\cal{D}_L$
takes place for adiabatic change of $\lambda_1$.
%, for a freezing interval, $l_f$, equal to $\lambda_2$. $l_f$ increases
%is equal to $\lambda_2$ 
%and the frozen QD value, $\cal{D}_L^f$ decreases with increasing $\lambda_2$. We observe that for $\lambda_2 \approx 0$ (black-diamond), $\cal{D}_L$ decays 
%in a manner similar to the isotropic case $\lambda_1 = \lambda_2$ (red-continuous). 
The behavior of entanglement is given in the inset. {All quantities are dimensionless, except QD (in bits) 
and entanglement (in ebits)}. Here, $N = 10$.
}
\label{fig:1}
\end{center}
%\vspace{-0.5cm}
\end{figure}

Let us now consider the case where $\cal{H}_{bulk}$ represents an ordered, \emph{XX} spin chain, such that $\cal{J}_i = \cal{K}_i = 1, \forall~ i = 2,...,N-2$, and the coupling strengths at the ends
% spins are different along the $X$- and the $Y$--axes, i.e., $\lambda_1 \neq \lambda_2$, 
are such that $\lambda_1, \lambda_2 < \cal{J}_i$. 
%We call this the 
%We refer to the situations 
When $\lambda_1\neq\lambda_2$, we refer to it as ``unbalanced'' weak-coupling. Under such coupling, the behavior of $\cal{E}_L$ and long-range QD ($\cal{D}_L$) i.e., the QD of the reduced density matrix of the end spins, $\rho_{1N}$, of the ground state, are qualitatively different. Specifically, for fixed values of $\lambda_2$ and on slowly increasing the parameter $\lambda_1$, from approximately $0$ to $1$, $\cal{D}_L$  freezes, i.e., remains unchanged in value
%, from $\lambda_1 \approx 0$ to $\lambda_1 = \lambda_2$, i.e., 
for the range 0 $<\lambda_1\le\lambda_2$. The value of the frozen $\cal{D}_L$ ($\cal{D}_L^f$) is dependent on the fixed $\lambda_2$ and the size of the spin chain, $N$. However, the freezing interval ($l_f$), i.e., the region on the $\lambda_1$-axis over which $\cal{D}_L$ remains frozen, is equal to $\lambda_2$. As $\lambda_1$ is increased beyond $\lambda_2$, i.e., for the range $\lambda_1 > \lambda_2$, the freezing of $\cal{D}_L$ ceases and it decays exactly similar to the balanced case. 
Interestingly, the behavior of entanglement 
does not distinguish the regimes 0 $<\lambda_1 \le \lambda_2$ and $\lambda_1 > \lambda_2$.
In particular, $\cal{E}_L$  decays in a similar fashion in both the regimes,
%%is 
%%the same in the two regimes,
though the maximum $\cal{E}_L$ at $\lambda_1 \approx 0$ and the 
value of $\lambda_1$ at  the non-temporal death (i.e., $\lambda_1^D$, such that for all $\lambda_1 \ge \lambda_1^D, \cal{E}_L = 0$), both decrease with increasing $\lambda_2$. 
Figure~\ref{fig:1} shows 
%this ``adiabatic freezing'', i.e., 
the freezing of long-range QD 
%in the adiabatic evolution obtained by slow variation of the weak coupling, $\lambda_1$, 
%of quantum discord 
in an $N = 10$ spin \emph{XX} chain for different values of the fixed weak coupling, $\lambda_2$, with the variation of $\lambda_1$. 
We call such phenomenon as ``adiabatic freezing'' since it can be observed in the adiabatic evolution obtained by slow variation of the weak coupling $\lambda_1$. The phenomenon of adiabatic freezing can be explained by observing the behavior of the two-site correlation functions that are obtained by solving the Hamiltonian in Eq.~(\ref{eq:ham3}). Note that the freezing of $\cal{D}_L$ observed in the present context is purely a property of the {quantum spin chain, and not a conjunction of the system and an environment. 
%See the Supplemental Material \cite{supple} for a more detailed exposition of the adiabatic freezing phenomenon.}
%%We observe that %the freezing length is equal to fixed weak coupling, i.e., $l_f$ = $\lambda_2$. 
%%%Interestingly, there exists a complementary behavior between $l_f$ and the frozen discord value, $\cal{D}_L^f$ (cf. \cite{titas}).
%%%For $\lambda_1 > \lambda_2$, the decay of $\cal{D}_L$ is identical to the balanced weak coupling case, $\lambda_1 = \lambda_2$, which is also equal to the behavior for $\lambda_2 \approx 0$. 
%The inset of Fig. \ref{fig:1} shows the behavior of entanglement for the same parameter regime. 

{We note that both $\cal{E}_L$ and $\cal{D}_L$ are quasi long-range \cite{venuti}, i.e., their value  decreases with increase in system-size $N$. Hence, for a fixed $\lambda_2$, the frozen discord value, $\cal{D}_L^f$, decreases as $N$ grows larger. However, fixing the weak-end couplings to low values, say $\lambda_1/\cal{J}_i \approx 0.001$, for $N$=200, a reasonably high $\cal{D}_L^f \approx$ 0.93 can be obtained, and frozen upto one order higher, i.e., $l_f=\lambda_2$ = 0.01. {Moreover}, even symmetric quantum correlation measures, such as symmetric QD \cite{sym-dis}, exhibits adiabatic freezing for $\lambda_1\neq\lambda_2$. }
%\cite{supple}.

%%%Let us now discuss the adiabatic freezing of quantum discord with increasing $N$. Figure \ref{fig:2} shows the behavior of quantum discord and entanglement, with respect to the weak coupling parameter $\lambda_1$, for a fixed value of $\lambda_2 (= 0.2)$, for different number of spins ($N$). The figure reveals that the freezing length, $l_f$, remains constant, equal to $\lambda_2$, as the number of spins is increased. However, the frozen discord value, $\cal{D}_L^f$, decreases with the increase in system size
%%%and hence we can see another dual nature, viz. between  $\cal{D}_L^f$ and $N$.
%%%\begin{figure}[t]
%%%\begin{center}
%%%\includegraphics[width=6cm]{fig2.eps}
%%%%\epsfig{figure= try2.eps}
%%%\caption{(Color online.) Behavior of adiabatic freezing with system-size. We plot the adiabatic dynamics of long-range quantum discord 
%%%and long-range entanglement (inset) 
%%%with increase in end-spin coupling $\lambda_1$, for a fixed, $\lambda_2 = 0.2$,  with $N = 10, 20, 50, 100, 200$ spins. 
%%%The  bulk spins are governed by the $XX$ model.
%%%We observe that the freezing length, $l_f$, remains constant, equal to $\lambda_2$, as the number of spins is increased but the frozen discord value, $\cal{D}_L^f$, decreases. The units are the same as in Fig. \ref{fig:1}.
%%%}
%%%\label{fig:2}
%%%\vspace{-0.6cm}
%%%\end{center}
%%%\end{figure}

\begin{figure}[h]
\begin{center}
\epsfig{figure= 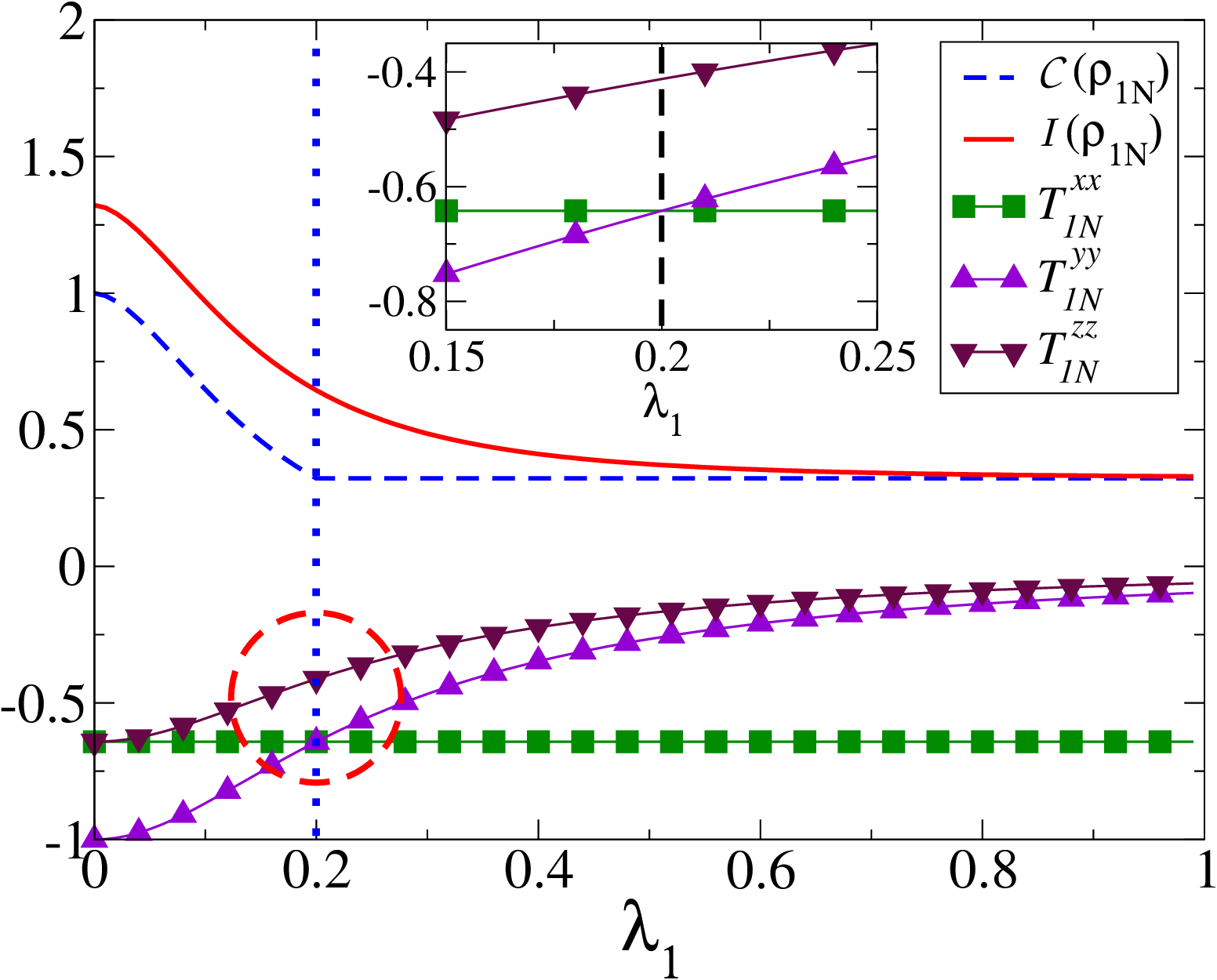,
%try3.eps, 
height=.19\textheight, width=0.35\textwidth,angle=-0}
 \caption{(Color online.) 
%Understanding adiabatic freezing. 
%Behavior of classical correlations and mutual information. 
%The figure shows the v
{Variation of $\cal{I}(\rho_{1N})$, $\cal{C}(\rho_{1N})$, and $T^{\alpha\alpha}_{1N}$ with increase in the end coupling $\lambda_1$, for a fixed $\lambda_2 = 0.2$, and 
for $N$ = 20 spins, where the bulk is an \textit{XX} chain. $\cal{D}_L$ is frozen for $\lambda_1$ $\le$ $\lambda_2$. $\cal{D}_L^f$ = 0.322 and $l_f$ = 0.2. 
All quantities used are dimensionless except  $\cal{D}_L$, $\cal{I}(\rho_{1N})$, and $\cal{C}(\rho_{1N})$, which are in bits. The inset magnifies the red-encircled region}. 
%The figure shows the dynamics of long-range quantum discord (red-squared) and long-range entanglement (blue-circled) with increase in the X-axis end coupling $\lambda_1$, for a fixed Y-axis weak coupling value, $\lambda_2 = 0.2$, for an open $XX$--chain with $N = 20$ spins. We observe that quantum discord is frozen for $\lambda_1 \le \lambda_2$, with $\cal{D}_L^f$ = 0.322, and $l_f$ = $\lambda_2$ = 0.2. The value of $\lambda_1$ for which ESD occurs is $\lambda_1^D$ = 0.59.
}
\label{fig:3}
\end{center}
%\vspace{-0.5cm}
\end{figure}

The analysis of two-site correlation functions between the end spins, $T^{\alpha\alpha}_{1N}$, 
%, obtained from Eq.~(\ref{corr}), 
%that contribute to the formation of the two-site end-to-end density matrix 
can shed light on the observed adiabatic freezing of QD.
%
%The observed freezing of quantum discord can be explained by analyzing the two-site correlation functions ($T^{\alpha\alpha}_{1N}$) that contribute to the formation of the two-site end-to-end density matrix. 
{From Fig.~\ref{fig:3}, for $N$ = 20 spins with $\lambda_2$ = 0.2, we observe that $T^{\alpha\alpha}_{1N} <$ 0, and $|T^{\alpha\alpha}_{1N}| <$ 1, $\forall~\alpha$. Since, $T^{zz}_{1N}$ = $-~T^{xx}_{1N}~T^{yy}_{1N}$ \cite{LSM,supple}, therefore $T^{xx}_{1N}$ and $T^{yy}_{1N}$ are the only independent variables, with $|T^{zz}_{1N}| \leq |T^{xx}_{1N}|$ and $|T^{zz}_{1N}| \leq |T^{yy}_{1N}|$. Moreover, 
%$T^{xx}_{1N}$ remains constant with the variation of $\lambda_1$.}
%
it is seen that $T^{xx}_{1N}$ is constant during the adiabatic evolution.
%, i.e., it is solely a function of the fixed weak end coupling, $\lambda_2$. This is evident from Fig.~\ref{fig:3}. 
%As $T^{zz}_{1N}$ is a function of  $T^{xx}_{1N}$ and  $T^{yy}_{1N}$,  and $T^{xx}_{1N}$ is constant, 
Hence, the quantum mutual information ($\cal{I}(\rho_{1N})$) is an entropic function of $1 \pm T^{yy}_{1N}$. Since, $|T^{yy}_{1N}| \ge |T^{xx}_{1N}| \ge |T^{zz}_{1N}|$ for $\lambda_1 \le \lambda_2$, the classical correlation ($\cal{C}(\rho_{1N})$) is also a function of $1 \pm T^{yy}_{1N}$. Hence, $\cal{I}(\rho_{1N})$ and $\cal{C}(\rho_{1N})$ decay with identical rates leading to a constant long-range QD. % ($\cal{D}_L(\rho_{1N})$ = $D_L^f$). 
However, for $\lambda_1 > \lambda_2$, $|T^{xx}_{1N}| > |T^{yy}_{1N}| \ge |T^{zz}_{1N}|$ and $\cal{C}(\rho_{1N})$ is now a function of $1 \pm T^{xx}_{1N}$. Since, $T^{xx}_{1N}$ is constant during the evolution, $\cal{C}(\rho_{1N})$ freezes for $\lambda_1 > \lambda_2$, in contrast to $\cal{I}(\rho_{1N})$ in that range. Hence, the freezing of the long-range $\cal{D}_L(\rho_{1N})$ disappears for $\lambda_1 > \lambda_2$. As $\lambda_2 \to \{0,\lambda_1\}$, there is no freezing of $\cal{D}_L(\rho_{1N})$, as $\cal{C}(\rho_{1N})$ is always constant.
In contrast, for entanglement, no adiabatic freezing occurs. In Fig.~\ref{fig:3}, since 0 $< |T^{\alpha\alpha}_{1N}| <$ 1, $\forall~\alpha$, $\cal{E}_L$ is given by the relation, $\cal{E}_L$ = $\max\left[0, 1/2(g_{1N} - h_{1N})\right]$, where, $g_{1N}$ = $|T^{xx}_{1N}+T^{yy}_{1N}|$, and $h_{1N}$ = $1-|T^{xx}_{1N}||T^{yy}_{1N}| >$ 0. As $\lambda_1$ increases $g_{1N}$ and $h_{1N}$ decreases, due to decreasing $|T^{yy}_{1N}|$, and constant $|T^{xx}_{1N}|$. For $g_{1N} > h_{1N}$,  $\cal{E}_L$ decreases with $\lambda_1$, and at $g_{1N} \leq h_{1N}$, $\cal{E}_L$ = 0 (sudden death). 
%The above analysis can be compared with known results on quasi long-range entanglement. 
For $\lambda_1$ = $\lambda_2$, we have $T^{xx}_{1N}$ = $T^{yy}_{1N}$ = $z$ (say). 
%Moreover, $T^{zz}_{1N}$ = $-z^2$, which gives us the relation for the long-range entanglement, 
Therefore, $\cal{E}_L$ = $\max\left[0, 1/2(z^2+2|z|-1))\right]$ \cite{supple}. }
%Now for $x$ = $-\langle S_1^x S_N^x + S_1^y S_N^y\rangle$ = $-1/2~ z$, we have $\cal{E}_L$ = $2\max\left[0, (x^2+|x|-1/4))\right]$, as shown in \cite{venuti}.}
%%, which in turn is due to the fact that the condition $|T^{xx}_{1N}| > |T^{yy}_{1N}| \ge |T^{zz}_{1N}|$ is always satisfied and $T^{xx}_{1N}$ is constant. The above conditions can be explicitly observed in Fig.~\ref{fig:3}, for a spin chain with $N$ = 20 spins. {The adiabatic evolution of $\cal{D}_L(\rho_{1N})$ and $\cal{E}_L(\rho_{1N})$ with respect to the correlation functions $T^{\alpha\alpha}_{1N}$ is discussed further in the supplementary material.}

The observed behavior of $\cal{I}(\rho_{1N})$ and $\cal{C}(\rho_{1N})$ is consistent to what is observed in the freezing of QD, first observed in dephasing quantum systems \cite{manis}, that heralded a substantial amount of research in recent years \cite{expt,freezing,titas}. 
%The phenomenon of adiabatic freezing can be explained by observing the behavior of the two-site correlation functions 
%\cite{supple} that are obtained by solving the Hamiltonian in Eq. (\ref{eq:ham3}).
However, we note that the freezing of $\cal{D}_L(\rho_{1N})$ observed in the present context is purely a property of the long-range correlations in the spin chain and is devoid of any external decoherence, in contrast to conventional freezing phenomena.

\section{Detecting anisotropy transition}\label{sec:DNS}
The adiabatic freezing of the quasi long-range QD ($\cal{D}_L$) in spin chains, with Hamiltonians of the form given by 
Eq.~(\ref{eq:ham3}), can be used to investigate certain intrinsic properties of the bulk Hamiltonian, such as the anisotropy transition.
{Though there exist methods to detect anisotropy in quantum spin systems \cite{anis-expt}, our results provide an information-theoretic perspective to investigate these quantum properties in an experimentally viable manner.}

% present in the bulk Hamiltonian. 
Let $\cal{H}_{bulk}$ be an ordered \emph{XY} spin chain with $\cal{J}_i$ = $1 + \gamma$ and $\cal{K}_i$ = $1 - \gamma$, $\forall~ i=2,...,N-2$, {with $\gamma(\neq 0)$ being the finite anisotropy} present in $\cal{H}_{bulk}$. The end-coupling is unbalanced, such that $\lambda_1 \neq \lambda_2$. {We focus on the parameter regime 
%Without any loss in generality, we consider the case, 
$\lambda_1, \lambda_2 \leq 1$, where the relevant physics under consideration is clearly observed}. 
%For $\gamma = 0$, we recover the model discussed in the previous section.
%
Let us now consider the {adiabatic freezing of $\cal{D}_L$ in the {above} spin chain, with possible anisotropy in $\cal{H}_{bulk}$}, by varying the weak end-couplings.
% along the $X$--axis ($\lambda_1$) and keeping the weak end-coupling along the $Y$--axis ($\lambda_2$) fixed. 
The {presence of} anisotropy
% in the bulk, 
introduces changes in the freezing characteristics of $\cal{D}_L$. {For instance,
%frozen quantum discord 
$\cal{D}_L^f$} increases with $\gamma$ upto a certain fixed value 
%of $\gamma$, 
that depends on $N$, and then decreases. See inset of Fig.~\ref{fig:4}.
\begin{figure}[h]
\begin{center}
\includegraphics[width=6cm]{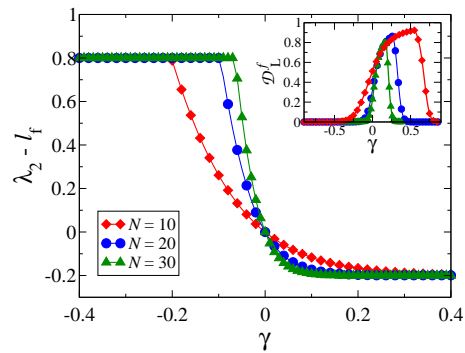}
\caption{(Color online.) Detecting the anisotropy ($\gamma$) transition. The order parameter, $\lambda_2 - l_f$ is plotted against $\gamma$, with fixed $\lambda_2$. 
%$l_f$ is the freezing interval during the variation of $\cal{D}_L$ with $\lambda_1$. We observe that $\lambda_2 - l_f$ vanishes for $\gamma$ = 0 and is finite otherwise. The sharp decrease of $\lambda_2 - l_f$ occurs at $\gamma^N_c$. For large but finite $N$, $\gamma^N_c \to 0$. 
The inset shows that $\cal{D}_L^f$ can be substantially increased for positive $\gamma$, below a threshold value. 
{The anisotropy transition can also be seen from the change in the curvature (convexity to concavity) of $\cal{D}_L^f$ with respect to $\gamma$, where the $N$-dependent critical values approach 
$\gamma = 0$ through negative $\gamma$.}
%, or from the sharp change in the rate of variation of $\cal{D}_L^f$ with respect to $\gamma$, where the critical values approach $\gamma = 0$ through positive $\gamma$. 
%In all the plots, red diamonds, blue circles, and green triangles represent the cases for $N =$ 10, 20, and 30 spins, respectively. 
All quantities are dimensionless, except QD (in bits).
%The figure shows the behavior of $\lambda_2 - l_f$ for a spin chain with $N = 10$ (red-diamond), and $N = 20$ spins (blue-circle).
} 
\label{fig:4}
\end{center}
%\vspace{-0.5cm}
\end{figure}
%a specific value. Under positive values of anisotropy ($\gamma > 0$), the $\cal{D}_L^f$ can be increased to $0.92$ for $N = 10$, and $0.87$ for $N = 20$ spins (see inset, Figure \ref{fig:4}). 
Interestingly, {the freezing interval,
%behavior of freezing length 
$l_f$}
%, under variation of $\gamma$, 
undergoes a transition that can detect the anisotropy transition at $\gamma = 0$ in the bulk spin system. This allows us to define an anisotropy transition order in terms of the system parameters and {the freezing interval $l_f$}. 
%If we consider, $|\lambda_2 - l_f|$ as a figure of merit, where $\lambda_2$ is the controlled weak end-spin coupling and the $l_f$ is characteristic of the model under consideration, then 
This can be seen as follows. 
{We evaluate $\l_f$ for different values of $\lambda_1$, and a fixed $\lambda_2$. 
%$l_f$ can therefore be seen as a function of $\lambda_2$
By using $\lambda_2 -l_f$ as the order parameter, one can detect the $\gamma$--transition in the system at $\gamma = 0$, driven by changes in the anisotropy parameter. The order parameter $\lambda_2 -l_f$ is finite for anisotropic systems and vanishes at $\gamma$ = 0.}
{Further, one observes that $\lambda_2 -l_f$ sharply decreases as $\gamma$ is increased beyond an $N$-dependent threshold value ($\gamma^N_c$).
% that coincides with the point where the long-range discord sharply increases. 
It is observed that as $N$ is increased upto certain values, the critical parameter $\gamma^N_c \to 0$. Hence, for large but finite $N$, transition in $\lambda_2 -l_f$ at $\gamma^N_c$ captures the 
$\gamma$--transition of the class of quantum spin models described by the Hamiltonian in Eq.~(\ref{eq:ham3}). %Further, the order parameter $\lambda_2 -l_f$ is finite for anisotropic systems and vanishes 
%at $\gamma$ = 0. 
The observed transition is shown and described in Fig.~\ref{fig:4}. }

In an experimental setting, one may consider the end spins to be probe sites at the edge of a quantum spin chain consisting of the bulk spins. {The two end-spins can be defects or scattering particles in a spin chain \cite{venuti,vencite}, and the weak end-coupling can be experimentally controlled}. Under such conditions, the anisotropy of $\cal{H}_{bulk}$ can be detected using the order parameter $\lambda_2 - l_f$, where the {desired quantities are observed from the freezing of long-range QD between the two probe spins}. Hence, the 
%freezing of long-range quantum discord between end spin sites of an ordered, open quantum spin chain, serving as probes, can detect the presence of anisotropy in the bulk spin chain and the 
quantity $\lambda_2 - l_f$ can serve as a suitable order parameter in many-body simulations.
The advantage of the approach lies in the fact that all performed measurements and 
tuning of interactions are associated only with the probe spins that weakly interact with the quantum spin chain. The detection of {the intrinsic} anisotropy in the bulk spin chain is achieved without disturbing the system. This provides an interesting role for adiabatic freezing of discord in investigating many-body phenomena.

\section{Freezing of energy gap}
{Another interesting phenomenon that connects adiabatic freezing of long-range quantum correlations to the cooperative properties of the quantum spin chain, is the freezing of energy gap.} 
{For the considered model, the weak end-couplings between the bulk and end spins introduces a finite energy gap.}

\begin{figure}[h]
\begin{center}
\includegraphics[width=6cm]{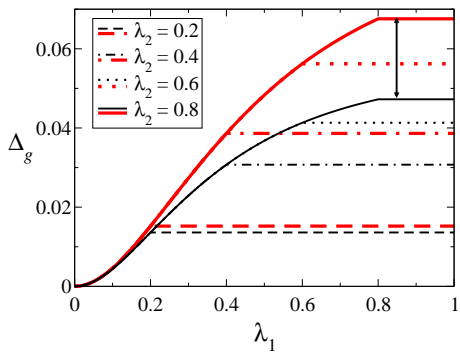}
\caption{(Color online.) Adiabatic freezing of the energy gap. {Numerically obtained values of $\Delta_g$ is plotted against $\lambda_1$, for 
different fixed values of $\lambda_2$. }
%The figure shows the change in energy gap ($\Delta_g$) with the end-coupling parameter, $\lambda_1$, for 
The spin model is the same as in Fig.~\ref{fig:1}, but for $N$ = 20 (red) and 30 (black). 
%
%We observe that $\Delta_g$ increases with $\lambda_1$, for $\lambda_1 < \lambda_2$. However, 
For $\lambda_1 \ge \lambda_2$, $\Delta_g$ is frozen.
% at a fixed value. The behavior of $\kappa \Delta_g$, 
%in terms of freezing, is similar to the classical correlation, $\cal{C}$ in Ref.~\cite{supple}. We observe that the frozen value of $\Delta_g$ is greater for higher $\lambda_2$, and it decreases with increasing $N$. 
The difference in frozen $\Delta_g$ for different $N$, 
%given by $\delta_{N'}$, increases with $\lambda_2$.
marked in one case with a double-headed arrow, increases with $\lambda_2$. All quantities plotted are dimensionless. Note that the behavior of $\Delta_g$ and $\cal{D}_L$ with response to $\lambda_1$, for fixed $\lambda_2$, is complementary.
} 
\label{fig:5}
\end{center}
%\vspace{-0.5cm}
\end{figure}

The energy gap {is an intrinsic property} of a quantum spin system that can be obtained from the excitation energy spectrum or the dispersion relation of the function $\Delta_k$ \cite{LSM,supple}. 
Let us consider the system, %discussed in Sec. \ref{freeze}, 
where $\cal{H}_{bulk}$ in Eq.~(\ref{eq:ham2}), represents an ordered \emph{XX} spin chain.
%, with $\cal{J}_i = \cal{K}_i = 1, \forall~ i =2,...,N-2$. 
The only anisotropy in the system arises from the coupling strengths at the end spins.
% along the $X$- and the $Y$--axes, i.e., $\lambda_1$ and $\lambda_2$, such that $\lambda_1, \lambda_2 \neq \left\{J_i\right\}$. 
{Solving the Hamiltonian,
%\cite{supple}}, 
one can obtain the excitation spectrum, $\Delta_k$. 
%as $n^{th}$ roots of some polynomial function. 
%by finding the roots of some polynomial of degree $N/2$. 

{For the balanced case, $\lambda_1$ = $\lambda_2$ = $\lambda$, the dispersion relation of the excitation energy is given by $\Delta_k$ = $\cos(k)$, where $k$ are the quasimomenta modes. These modes satisfy the eigenvalue equation \cite{grudka},
$
\mu \cot(k)[\cot((N-1)k/2)]^\mu = \lambda^2/(2-\lambda^2)
$, for $\lambda\neq$ 1, with $\mu$ = $\pm1$ being the eigenstate parity. Considering $\mu$ = 1,
the energy gap is then given by $k'$, which minimizes $\Delta_{k'}$ = $\cos(k')$, while satisfying the above eigenvalue equation.  
For $N \gg$ 1, at $k' = \pi/2-\delta$, where $\delta\rightarrow 0$, an analytical expression for the energy gap ($\Delta_g$) is obtained \cite{venuti}, where
$
\Delta_g \approx ({\pi}/{2N})[(1 + {2}/({N(\lambda^2/(2-\lambda^2)+2})] = f(\lambda).
$
Numerical analysis shows that for the case, $\lambda_1 \neq \lambda_2$, the quasimomenta $k'$ corresponding to the energy gap satisfies the eigenvalue equation for $\lambda$ = $\min[\lambda_1,\lambda_2]$. 
Therefore, the energy gap is given by the relation, $\Delta_g = \min[f(\lambda_1), f(\lambda_2)]$, where $f(\lambda)$, defined above, is a monotonically decreasing function of $\lambda$.  
The adiabatic freezing of the numerically estimated energy gap, is shown in Fig.~\ref{fig:5}. Using the above relation for $\Delta_g$ and monotonicity of the function $f(\lambda)$, it is obvious that $\Delta_g$ freezes for $\lambda_2 \leq \lambda_1$, since in this region, $f(\lambda_2)\leq f(\lambda_1)$ and $\Delta_g$ is thus independent of $\lambda_1$.

\begin{figure}[h]
\begin{center}
\includegraphics[width=6cm]{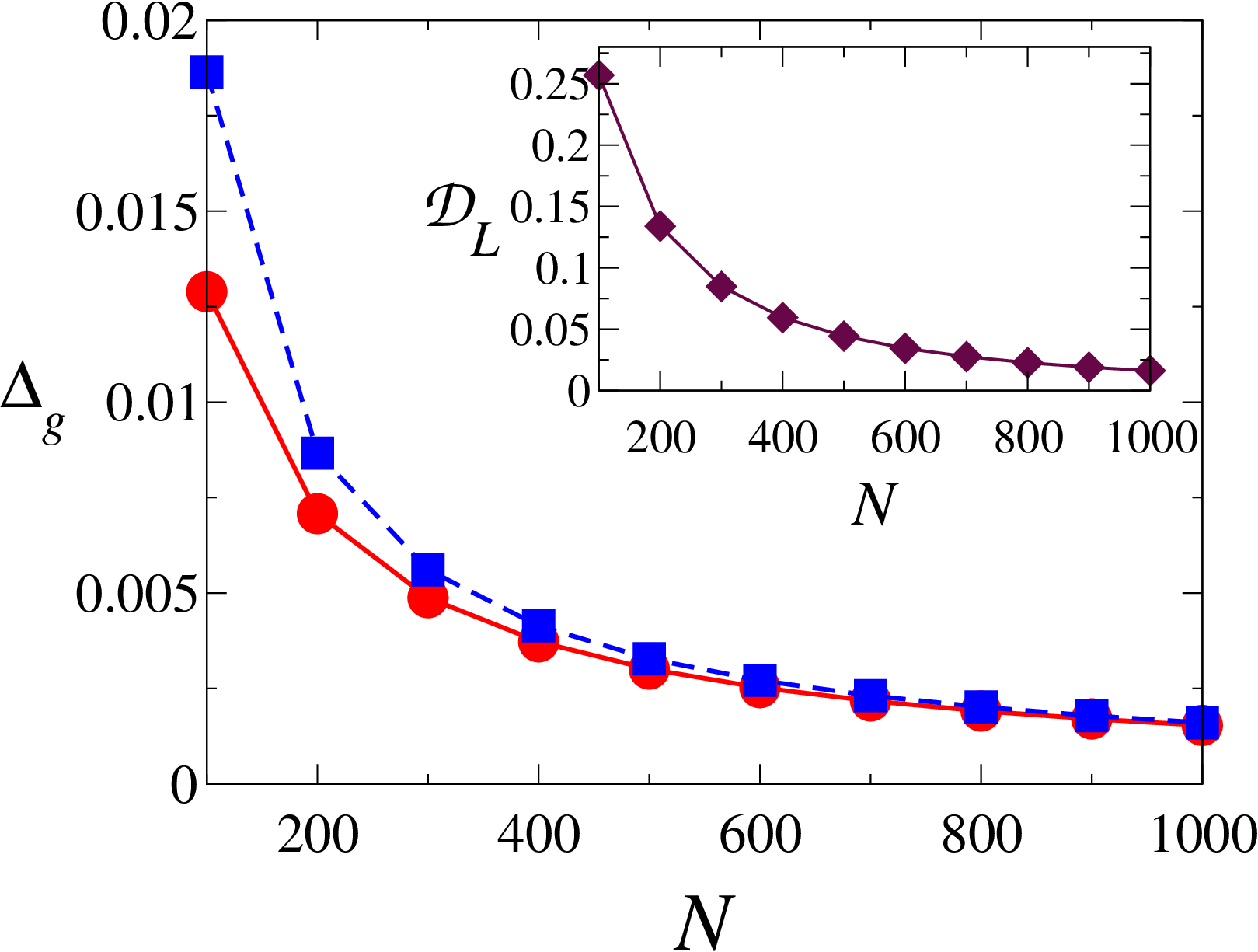}
\caption{(Color online.) {Variation of the analytically and numerically estimated values of the energy gap with increasing size of the spin chain. 
The analytical expression is given by $\Delta_g = \min[f(\lambda_1), f(\lambda_2)]$ (blue-square), and is observed to be consistent with numerically obtained values of $\Delta_g$ (red-circle) at large $N$, for $\lambda_1$ = 0.4 and $\lambda_2$ = 0.6 (and, $\lambda_1$ = 0.6 and $\lambda_2$ = 0.4). The inset shows $\cal{D}_L$ (maroon-diamonds) with increasing $N$. The figure shows that at large $N$, $\Delta_g$ and $\cal{D}_L$ scale as $1/N$.}
} 
\label{fig:5.5}
\end{center}
%\vspace{-0.6cm}
\end{figure}

The adiabatic freezing of $\Delta_g$ is complementary to that of $\cal{D}_L$, in terms of the variation of weak end-couplings. Figs.~\ref{fig:1} and \ref{fig:5} show that 
while $\cal{D}_L$ freezes for $\lambda_1 \leq \lambda_2$, $\Delta_g$ is constant for $\lambda_1 \geq \lambda_2$. Moreover, while the frozen value of $\cal{D}_L$ decreases with increase in $\lambda_2$, it does the opposite for $\Delta_g$. 
%In the supplementary material, we examine the validity of the analytical expression for $\Delta_g$ and compare the scaling and variations of $\Delta_g$ and $\cal{D}_L$, for large $N$. 
Figure~\ref{fig:5.5} shows the agreement between the analytically and numerically obtained values of $\Delta_g$, at large $N$. We observe that at large $N$, the frozen values of both $\Delta_g$ and $\cal{D}_L$ scale with $1/N$ \cite{supple}. 
{The behavior of $\Delta_g$ shows that the phenomenon of adiabatic freezing is manifested through the intrinsic properties of the quantum spin chain.}} %Recent experimental studies have shown that energy gap can be investigated using probe spins \cite{egap-probe}.}

%{} 

\section{Response to thermal fluctuations}
In the previous sections, the phenomena of freezing is observed for the ground state of the quantum spin chain under consideration. 
For small thermal fluctuations, the system is no longer in the ground-state, but a mixed state in equilibrium at some small temperature ($T$). 
To obtain the reduced two-site density matrix of the thermal equilibrium state of the spin system, at temperature $T$, one must find the thermal two-site correlation functions $T^{\alpha\alpha}_{ij}(\beta)$ \cite{LSM, supple}. 
%{The derivation of $\cal{G}_{ij}(\beta)$ is presented in the supplementary material}. 
Here, $\beta = 1/k_BT$, where $k_B$ is the Boltzmann constant.
\begin{figure}[h]
\begin{center}
\includegraphics[width=6cm]{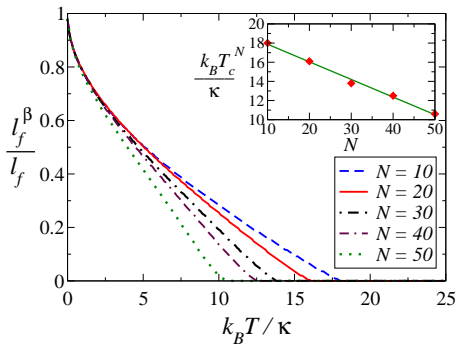}
\caption{(Color online.) Response of freezing to thermal fluctuations. We plot the ratio of the thermal freezing interval ($l_f^\beta$) and the ground state freezing interval ($l_f$) as a function of temperature. %,for quantum spin chains of sizes, $N$ = 10, 20, 30, 40, 50. 
%The ground state freezing length ($l_f$) is controlled by the weak end-coupling,
Here, $\lambda_2$= 0.2.
%The spin model considered here is the same as in Fig. \ref{fig:1}. 
We observe that for every $N$, there exists a critical temperature ($T^N_c$) beyond which adiabatic freezing disappears for finite chains.
%The inset shows the variation of $T^N_c$ with $N$. %We observe that 
$T^N_c$ decreases linearly with $N$, as shown in the inset. The solid green line, in the inset, is the linear fit, $y = 19.72 - 0.184 x$.
All quantities plotted are dimensionless.%, except $N$, which is the number of spins in the chain. 
The abscissae in the figure and the ordinates in the inset are multiplied by $10^4$.% for the ease of viewing.
} 
\label{fig:6}
\end{center}
%\vspace{-0.6cm}
\end{figure}

{  
%An important question that arises is whether the freezing is observed at finite $T$. 
We observe that adiabatic freezing persists at finite $T$, below a critical temperature $T^N_c$, that also depends on the system-size $N$. The value of frozen discord $\cal{D}_L^f$ is constant for temperatures below $T^N_c$, though the freezing interval decreases as $T$ increases. $T^N_c$ is a linearly decreasing function of $N$. Figure~\ref{fig:6} shows the effect of thermal excitations on freezing interval of long-range QD {in an \emph{XX} spin chain.}

\section{Discussion}
{In recent years, experimental generation of long-range quantum correlation in antiferromagnetic spins chains have been reported using strontium-cuprate compounds, such as Sr$_{14}$Cu$_{24}$O$_{41}$, to simulate dimerized spin-chains  \cite{new}. Quantum correlations are experimentally measured using low temperature magnetization, magnetic susceptibility, and heat capacity \cite{new,expt3}.
Another experimental protocol, relevant to the quantum spin chain considered in our study, was proposed  using ions of the ytterbium isotope, $^{171}$Yb, in segmented linear Paul traps \cite{iontrap}, utilizing the tailoring of axial trapping potential to generate the spin-spin coupling, and using microwave pulses to generate effective spin interactions. 
A more recent study devises a similar scheme using superconducting flux qubits, using dc currents and microwave pulses to control the spin-spin interaction \cite{flux}.} {Moreover, recent experimental studies have also observed the freezing phenomena in various quantum systems \cite{expt}.}

%From an experimental perspective, the importance of our results on adiabatic freezing of long-range QD is striking. 

In our work, we find that there exists adiabatic freezing of quasi long-range quantum correlations in finite quantum spin chains. We show that the
observed phenomena is robust to the weak end-spin couplings and finite thermal fluctuations, which are the fundamental elements in experimental control of quantum systems. Further, by tuning the end-spin coupling one can obtain relatively high values of long-range QD. 
{Interestingly, we have observed that a finite interaction between the two end spins can vastly increase the shared QD between the sites, without affecting the freezing interval. This is intuitively plausible, as the finite end-to-end coupling encourages greater correlation between the end spins. 
%However the quantum correlation is then no longer a long-range effect.
% but nonetheless can be easily  implemented experimentally using control pulses. 
{Alternatively, one can also study the phenomena in the \emph{XX} and \emph{XY} spin chains in a transverse magnetic field. 
Preliminary investigations reveal that, for weak end-couplings, an effective freezing \cite{titas} of long-range QD can be characterized. 
For quantum spin chains such as the frustrated spin-1/2 $J_1-J_2$ model, spin-1/2 \textit{XX} chain with alternating interactions, and the spin-1 AKLT model, long-range quantum correlations are observed \cite{venuti}, although adiabatic freezing is absent.}
We also note that the phenomenon of adiabatic freezing can be utilized to experimentally detect properties of quantum spin-baths, 
modelled by an interacting bulk spin Hamiltonian, and probed by weakly interacting spins at ends of the bath. 

{To conclude,  we find the phenomenon of adiabatic freezing of quasi long-range QD 
%is a phenomena hitherto not observed 
in the closed dynamics of many-body quantum systems. Our work makes a connection between the temporal freezing of correlations, observed only in damped quantum systems, to the feature of long-range correlations in quantum spin chains. However, in contrast to temporal freezing, the adiabatic phenomena is an intrinsic property of the considered spin system. It has the ability 
to detect important cooperative phenomena in quantum spin models and in particular serve as an order parameter
for detecting the anisotropy transition in quantum XY models.
We 
%observe that the freezing is robust under finite thermal fluctuations and 
note that the phenomena is also observed for other 
%measures of quantum correlation 
system properties such as energy gap, and other quantum correlation measures such as symmetric discord \cite{sym-dis} and one-way quantum work-deficit \cite{QWD}.}

{With unprecedented developments to simulate quantum spin chains in different physical substrates and experimental techniques to characterize quantum correlations, the phenomenon of adiabatic freezing allows the generation of robust  long-range quantum correlations between distant parties, for application in future quantum technologies. }

%\MakeUppercase{Done}
%\vspace{-1cm}

% 
\clearpage
\newpage
%\part{supple}
%\pagebreak
%\title{Supplemental Material: Adiabatic freezing of long-range quantum correlations in spin chains}
\maketitle
%\begin{widetext}
%\vspace{10cm}
\begin{center}
\textbf{{\Large{Supplementary Material}}}
\end{center}
%\end{widetext}
%%%%%%%%%% Merge with supplemental materials %%%%%%%%%%
%%%%%%%%%% Prefix a "S" to all equations, figures, tables and reset the counter %%%%%%%%%%
\setcounter{section}{0}
\setcounter{equation}{0}
\setcounter{figure}{0}
\setcounter{table}{0}
\setcounter{page}{1}
\makeatletter
\renewcommand{\thesection}{S\arabic{section}}
\renewcommand{\theequation}{s\arabic{equation}}
\renewcommand{\thefigure}{s\arabic{figure}}
\renewcommand{\bibnumfmt}[1]{[S#1]}
\renewcommand{\citenumfont}[1]{S#1}

%\section*{Supplementary Material}

\section{Spin Hamiltonian and correlation functions}  
\label{method}

Let us consider an anistropic \textit{XY} quantum spin chain containing $N$ spins with a closed end. The Hamiltonian for such a system can be written as
\begin{equation}
\cal{H} = \sum_i^N \frac{\kappa}{4}~(\cal{J}_i~ \sigma^x_i \sigma^x_{i+1} + \cal{K}_i ~\sigma^y_i \sigma^y_{i+1}), 
\label{SHam}
\end{equation}
where $\sigma^{x(y)}_{N+1}$ = $\sigma^{x(y)}_{1}$.
%, satisfies the desired boundary condition. 
%
$\cal{J}_i$ and $\cal{K}_i$ are the dimensionless interaction strengths.
$\kappa$($> 0$) has the unit of energy. 
% between the nearest-neighbors $i$ and $i+1$, along the $X$-- ($Y$--) direction. 
$\sigma^{i}$, $i = x, y, z$, are the Pauli spin matrices.
The open-end case is obtained by setting $\cal{J}_N$ and $\cal{K}_N$ equal to zero. 

The two-site quantum correlation, between arbitrary sites, in the ground state of the Hamiltonian can be obtained by deriving the two-site reduced density matrix, following the seminal work in \cite{SLSM}. The Hamiltonian given in Eq.~(\ref{SHam}), can be transformed in terms of spin-raising and -lowering operators,
$\hat{a}^\dag_i$ =  $\sigma^{x}_i$ + $i\sigma^{y}_i$ and $\hat{a}_i$ =  $\sigma^{x}_i - i\sigma^{y}_i$, to obtain
\begin{equation}
\cal{H} = \frac{\kappa}{2}\sum_i (\cal{J}'_i~\hat{a}^\dag_i \hat{a}_{i+1} + \cal{K}'_i ~ \hat{a}^\dag_i \hat{a}^\dag_{i+1} +  \mathrm{h.c.}),
\end{equation}
where $\cal{J}'_i$ = $(\cal{J}_i$ + $\cal{K}_i)/2$ and $\cal{K}'_i$ = $(\cal{J}_i - \cal{K}_i)/2$. The partly-Fermi, partly-Bose operators ($\hat{a}^\dag$) can be transformed to a set of strictly Fermi operators ($\hat{k}^\dag$), using the Jordan-Wigner transformations \cite{SJW}, such that $\hat{k}_i$ = $\exp\left[i\pi \sum_{j=1}^{i-1} \hat{a}^\dag_j \hat{a}_j\right]\hat{a}^\dag_i$ and $\hat{k}_i^\dag $ = $\hat{a}^\dag_i
\exp\left[-i\pi \sum_{j=1}^{i-1} \hat{a}^\dag_j \hat{a}_j\right]$. 
Equation~(\ref{SHam}) takes  a quadratic form 
%of the Hamiltonian, 
in terms of the Fermi creation ($\hat{k}^\dag$) and annihilation ($\hat{k}$) operators, that can be diagonalized.
%, as shown next. 
The quadratic-form Hamiltonian is given by 
$
\cal{H}$ = $\kappa\sum_{ij} \hat{k}^\dag_i \cal{A}_{ij} \hat{k}_{j}$ + $\frac{1}{2}(\hat{k}^\dag_i  \cal{B}_{ij} ~ \hat{k}^\dag_{j} $ + $\mathrm{h.c.})
$,
where $\cal{A}_{ij}$ = $\frac{1}{2}(\cal{J}'_i \delta_{i+1,j}$ + $\cal{J}'_j \delta_{i,j+1})$ is a symmetric matrix and $\cal{B}_{ij}$ = $\frac{1}{2}(\cal{K}'_i \delta_{i+1,j}$ - $\cal{K}'_j \delta_{i,j+1})$ is an anti-symmetric matrix. For closed-ended chains, $\cal{A}_{1N}$ = $\cal{A}_{N1}$ = $\cal{J}'_N$ and $\cal{B}_{1N}$ = $-\cal{B}_{N1}$ = $\cal{K}'_N$. As shown in \cite{SLSM}, any two-site reduced density matrix of the ground state, for arbitrary sites, can be derived in terms of the matrices $\mathbf{\cal{A}}$ and $\mathbf{\cal{B}}$, by solving the eigenvalue equation, $\phi_k (\cal{A}$ - $\cal{B})(\cal{A}$ + $\cal{B})$ = $\Delta_k^2 \phi_k$. The dispersion relation of the function $\Delta_k$ gives us the excitation spectrum that can be used to estimate the energy gap in the system.
A corresponding vector, $\psi_k$, is defined as $\psi_k$ = $\frac{1}{\Delta_k} (\cal{A} + \cal{B}) \phi_k$. A unitary  correlation matrix, $\cal{G}$ is then obtained by the relation, $\cal{G}_{ij}$ = $-\sum_k \psi_{ki}~\phi_{kj}$.
%

%\vspace{1cm}
\begin{figure}[h]
%\begin{center}
\epsfig{figure= 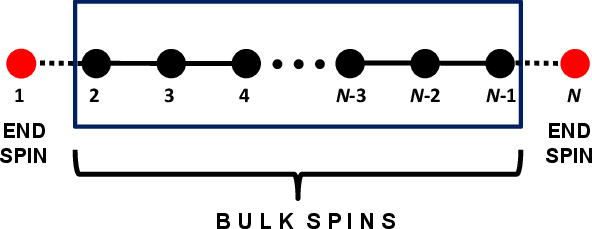,
%try3.eps, 
height=.12\textheight, width=0.42\textwidth,angle=0}
\caption{(Color online.) The set-up. The end spins are weakly coupled (black-dashed lines) with the bulk,
%in the quantum spin chain with $N$ spins. 
%The ends spins (red circle) are weakly coupled (black-dashed lines) to the bulk. 
which are strongly coupled to each other (black solid line). %The frozen discord is observed between spins $1$ and $N$. 
} 
\label{fig:0}
%\end{center}
\end{figure}

The two-site reduced density matrix can be obtained from the single-site magnetizations, $\langle \sigma^\alpha_i \rangle$, and the two-site correlation functions, $\langle \sigma^\alpha_i \sigma^\beta_j \rangle - \langle \sigma^\alpha_i \rangle\langle\sigma^\beta_j \rangle$, where $(\alpha,\beta = x,y,z)$. The symmetry of the Hamiltonian ensures that the only non-vanishing (single-site) magnetization is $\langle \sigma^z_i\rangle$. However, for no external fields, the single-site magnetization also vanishes. Moreover, the only non-vanishing two-site terms are $\langle \sigma^x_i \sigma^x_j \rangle$, $\langle \sigma^y_i \sigma^y_j \rangle$, and $\langle \sigma^z_i \sigma^z_j \rangle$, which gives us the two-site correlation functions, since $\langle \sigma^\alpha_i \rangle$ = 0, $\forall~\alpha$. 
In terms of the correlation matrix, $\mathcal{G}$, derived in the main text by diagonalizing the Hamiltonian in Eq.~(\ref{SHam}), the correlation functions are given by  \cite{SLSM},
\begin{eqnarray}
&&T^{xx}_{ij} =  \left| \begin{array}{ccc}
		 \cal{G}_{i,i+1} & ..& \cal{G}_{i,j}\nonumber\\
		 : &  &: \nonumber\\
		 \cal{G}_{j-1,i+1}&..& \cal{G}_{j-1,j}
		 \end{array}\right|,\\\nonumber\\
&&T^{yy}_{ij} =  \left| \begin{array}{ccc}
         \cal{G}_{i+1,i} & ..& \cal{G}_{i+1,j-1}\nonumber\\
		 : &  &: \nonumber\\
		 \cal{G}_{j,i}&..& \cal{G}_{j,j-1}
		 \end{array}\right|,\\\nonumber\\
&&T^{zz}_{ij} = ~(\cal{G}_{i,i}\cal{G}_{j,j} -~ \cal{G}_{i,j}\cal{G}_{j,i}).
\label{Acorr}
\end{eqnarray}
Note that $T^{xx}$ and $T^{yy}$ are minors of the determinant of $\cal{G}$. For nearest neighbors, the above correlations reduce to $T^{xx}_{i,i+1} = \cal{G}_{i,i+1}$, $T^{yy}_{i,i+1} = \cal{G}_{i+1,i}$, and $T^{zz}_{i,i+1} = -\cal{G}_{i,i+1}~\cal{G}_{i+1,i}$, since $\langle \sigma^z_i\rangle$ = $-\cal{G}_{ii}$ = 0.
For the end-to-end spin (see Fig.~\ref{fig:0}), two-site correlation function, ($T^{\alpha\alpha}_{1,N}$), the minor is an $N-1 \times N-1$ matrix and the expressions in Eq.~(\ref{Acorr}) can be simplified to obtain the following: 
\begin{eqnarray}
T^{xx}_{1,N} &=& -\cal{G}_{N,1}\det{(\cal{A}-\cal{B})}/|\det(\cal{A}-\cal{B})|= \cal{G}_{N,1}, \nonumber\\
T^{yy}_{1,N} &=& -\cal{G}_{1,N}\det{(\cal{A}-\cal{B})}/|\det(\cal{A}-\cal{B})|=\cal{G}_{1,N}, \mathrm{and}\nonumber\\
T^{zz}_{1,N} &=& -\cal{G}_{1,N}~\cal{G}_{N,1} = - T^{xx}_{1,N}~ T^{yy}_{1,N}.
\label{Acorr-lr}
\end{eqnarray}

Now, any two-site reduced density matrix, for arbitrary sites $i$ and $j$, can be written as
\begin{equation}
\rho_{ij} = \frac{1}{4}(\mathbb{I} + \sum_{\alpha = x,y,z} T^{\alpha\alpha}_{ij}~ \sigma^\alpha_i \otimes \sigma^\alpha_j),
\label{Arho12}
\end{equation}
where $\mathbb{I}$ is the two-qubit identity matrix. Since the derivation of the two-site density matrix depends on the diagonalization of $N \times N$ matrices, such as $\cal{A}$ and $\cal{B}$, the method can be executed for chains with a large number of spins, and allows us to study the asymptotic behavior of several system properties. In our case, these are the long-range quantum correlations.

To obtain the reduced two-site density matrix of the thermal equilibrium state of the quantum spin system, at temperature $T$,  one must find the thermal correlation matrix $\cal{G}_{ij}(\beta)$, in a similar fashion to the analytical derivation done above. Here, $\beta = 1/k_BT$, where $k_B$ is the Boltzmann constant. The correlation matrix can be written as 
\begin{equation}
\cal{G}_{ij}(\beta) = -\sum_{k} \psi_{ki}~\phi_{kj} (\langle \eta_k^\dag \eta_k\rangle_\beta - \langle\eta_k \eta_k^\dag\rangle_\beta), 
\end{equation}
where $\eta_k^\dag$ are the spinless fermionic operators that diagonalize the Hamiltonian,
%in Sec. \ref{method}, 
thus generating fermionic excitations in the ground states with energy $|\Delta_k|$. 
From Fermi statistics, $\langle\eta_k \eta_k^\dag\rangle_\beta = 1/(\exp[\beta\Delta_k] +1)$, and hence, $\cal{G}_{ij}(\beta) = -\sum_{k} \psi_{ki} \tanh[\beta\Delta_k/2]\phi_{kj}$. Using $\cal{G}_{ij}(\beta)$, the reduced two-site density matrix for the thermal equilibrium state can then be evaluated by following the expressions for the ground state of the spin system as shown in Eqs.~(\ref{Acorr} - \ref{Acorr-lr}).

\section{Measures of quantum correlation}

The correlation functions $T^{\alpha\alpha}_{ij}$ can be used to derive quantum correlation measures, such as quantum discord and entanglement, for reduced two-site density matrices of both the ground and thermal equilibrium states of the system.
The obtained two-site reduced density matrix, in Eq.~(\ref{Arho12}), is Bell-diagonal and hence its quantum discord \cite{Sdisc} can be calculated using an analytical optimization \cite{SLUO}. For the Bell-diagonal density matrix, $\rho_{ij}$, its eigenvalues, $e_i$, can be obtained in terms of $T^{\alpha\alpha}_{ij}$:
\begin{eqnarray}
e_1 &=& 1/4(1 - T^{xx}_{ij} - T^{yy}_{ij} - T^{zz}_{ij});\nonumber\\
e_2 &=& 1/4(1 - T^{xx}_{ij} + T^{yy}_{ij} + T^{zz}_{ij});\nonumber\\
e_3 &=& 1/4(1 + T^{xx}_{ij} - T^{yy}_{ij} + T^{zz}_{ij});\nonumber\\
e_4 &=& 1/4(1 + T^{xx}_{ij} + T^{yy}_{ij} - T^{zz}_{ij}).
\label{Aevs}
\end{eqnarray}
%%$
%%e_1 = 1/4 - T^{xx}_{ij} - T^{yy}_{ij} - T^{zz}_{ij};
%%e_2 = 1/4 - T^{xx}_{ij} + T^{yy}_{ij} + T^{zz}_{ij};
%%e_3 = 1/4 + T^{xx}_{ij} - T^{yy}_{ij} + T^{zz}_{ij};
%%e_4 = 1/4 + T^{xx}_{ij} + T^{yy}_{ij} - T^{zz}_{ij}.
%%$
The quantum mutual information is given by the relation, $\cal{I}(\rho_{ij})$ = $\sum_i e_i \log_2 (4e_i)$. The classical correlation obtained after optimization over measurements on a single-party, is given by the relation
\begin{eqnarray}
\cal{C}(\rho_{ij}) &=& \sum_{k=1}^2 x_k \log_2 (2x_k), ~~\textrm{where}\\
%\end{eqnarray}
%where 
x_k &=& (1 + (-1)^kx)/2,~~ \textrm{for}~~ k = (1,2),~~ \textrm{and}\\
x &=&  \max\{\lvert T^{xx}_{ij}\rvert,\lvert T^{yy}_{ij}\rvert,\lvert T^{zz}_{ij}\rvert\}.
\end{eqnarray}
The quantum discord
%$\mathcal{D}(\rho_{ij})$ = $\cal{I}(\rho_{ij}) - \cal{C}(\rho_{ij})$, 
is then given by the relation, %\nonumber\\
\begin{eqnarray}
\mathcal{D}(\rho_{ij}) &=& \cal{I}(\rho_{ij}) - \cal{C}(\rho_{ij})\nonumber\\
&=& \sum_{i=1}^4 e_i \log_2 (4e_i) - \sum_{k=1}^2 x_k \log_2 (2x_k).
\label{Adisc}
\end{eqnarray}

Similarly, using concurrence \cite{Swoot} as our measure of choice, the entanglement between any two sites can be analytically derived. Concurrence of a two-qubit density matrix, $\rho_{ij}$, is defined by the relation
\begin{equation}
\cal{E}(\rho_{ij}) = \max\left[0,c_1-c_2-c_3-c_4\right],
\end{equation}
where $c_i$'s are the square root of the eigenvalues of the matrix $\rho_{ij}\tilde{\rho_{ij}}$, arranged in decreasing order.  
$\tilde{\rho_{ij}}$ = $\sigma^y_i\otimes\sigma^y_j~ \rho_{ij}^*~ \sigma^y_i\otimes\sigma^y_j$.
For the two-site density matrix obtained in Eq.~(\ref{Arho12}), $\tilde{\rho_{ij}}$ = ${\rho_{ij}}$, and $c_i$'s are nothing but the eigenvalues of ${\rho_{ij}}$, given by Eq.~(\ref{Aevs}), arranged in decreasing order. Hence, the concurrence of the obtained two-site reduced density matrix is given by
 %concurrence \cite{woot}. For the 
%Bell-diagonal density matrix, $\rho_{ij}$, the concurrence can be obtained using 
%the two-site correlation functions and is given by 
\begin{eqnarray}
\cal{E}(\rho_{ij}) 
&=& \max\left[0, 2~ e_{max} -1\right]\nonumber\\
&=& \max\left[0, \frac{1}{2}(|g^{+}_{ij}| - h^{+}_{ij}), \frac{1}{2}(|g^{-}_{ij}| -  h^{-}_{ij})\right],
\label{Aent}
\end{eqnarray}
where 
$e_{max}$ = $\max[\{e_i\}_{i=1}^4]$, 
$g^{\pm}_{ij} = T^{xx}_{ij} \pm T^{yy}_{ij}$, and $h^{\pm}_{ij}$ = $1~ \pm~ T^{zz}_{ij}$. Hence, once the correlation functions, $T^{\alpha\alpha}_{ij}$, are known from Eq.~(\ref{Acorr}), quantum discord and entanglement can be obtained using Eqs.~(\ref{Aent}) and (\ref{Adisc}), respectively.

To highlight the role of the correlation functions in the behavior of quantum discord and entanglement during the phenomena of adiabatic freezing, we consider an explicit example. Let us study the model considered in the main text: an $N$-spin open quantum spin chain, with nearest neighbor interactions, with two spins at the edge of the chain (end spins) are weakly coupled to the remaining bulk of $N-2$ spins, as shown in Fig.~\ref{fig:0}.
% through nearest neighbor interactions. 
As presented in the main text, the Hamiltonian for such a spin chain is given by, %$\cal{H}$ = $\cal{H}_{bulk} + \cal{H}_{end}$, where
\begin{eqnarray}
\cal{H} &=& \cal{H}_{bulk} + \cal{H}_{end},~~\textrm{where} \nonumber\\
\cal{H}_{{bulk}} &=& \sum_{i=2}^{N-2} \frac{\kappa}{4}~ (\cal{J}_i ~\sigma^x_i \sigma^x_{i+1} + \cal{K}_i~\sigma^y_i \sigma^y_{i+1}),\\ 
\cal{H}_{{end}} &=& \frac{\kappa}{4} \left[\lambda_1 (\sigma^x_{1} \sigma^x_{2} + \sigma^x_{N-1} \sigma^x_{N})\right.
\nonumber\\
&+& 
\left.\lambda_2 (\sigma^y_{1} \sigma^y_{2} + \sigma^y_{N-1} \sigma^y_{N})\right].
\label{Aeq:ham3}
\end{eqnarray}
$\lambda_1$ and $\lambda_2$ are the weak end-couplings, and $\left\{\cal{J}_i\right\}$ = $\left\{\cal{K}_i\right\}$ = 1, such that the bulk forms an XX spin chain. Adiabatic freezing of long-range quantum correlations is observed, when one of the end-couplings (say, $\lambda_2$) is kept fixed, while the other (say, $\lambda_1$) is adiabatically varied.

\begin{figure}[h]
%\begin{center}
%\vspace{0.5cm}
\epsfig{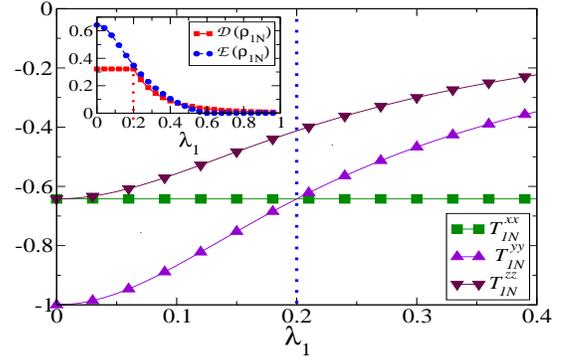}
 \caption{(Color online.) 
%Understanding adiabatic freezing. 
Variation of long-range two-site correlation functions, $T^{xx}_{1N}$ (green-square), $T^{yy}_{1N}$ (violet-up-triangle), and $T^{zz}_{1N}$ (maroon-down-triangle), with end-coupling strength $\lambda_1$, for a spin chain with $N$ = 20. The other end-coupling, $\lambda_2$ is kept fixed at 0.2. The inset figure shows the behavior of long-range concurrence (blue-diamond) and quantum discord (square-red). The adiabatic freezing of $\mathcal{D}_L$ for $\lambda_1 \leq \lambda_2$ is evident from the inset figure, whereas entanglement exhibits non-temporal death at $\lambda_1$ = 0.59. All quantities used are dimensionless except entanglement (in ebits) and quantum discord (in bits). Compare with Fig.~2 in the main text.
}
\label{vary}
%\end{center}
\end{figure}

In Fig.~\ref{vary}, one observes the adiabatic freezing of long-range quantum discord ($\mathcal{D}_L$) between the end spins, in the ground state of the Hamiltonian defined in Eq.~(\ref{Aeq:ham3}), when the end spin coupling satisfies the condition, $\lambda_1 \leq \lambda_2$ (fixed), $\forall~\lambda_1$. For $\lambda_1 > \lambda_2$ (fixed), $\mathcal{D}_L$ decays with increasing $\lambda_1$. This phenomena is however not observed for long-range entanglement ($\mathcal{E}_L$). The behavior of both $\mathcal{D}_L$ and $\mathcal{E}_L$ upon adiabatically varying $\lambda_1$ can be explained through the variation of the correlation functions $T^{xx}_{1N}$, $T^{yy}_{1N}$, and $T^{zz}_{1N}$ shown in Fig.~\ref{vary}.
We observe that $T^{\alpha\alpha}_{1N} <$ 0, with $|T^{\alpha\alpha}_{1N}| <$ 1, $\forall~\alpha = (x,y,z)$, for a spin chain with $N$ = 20 spins and $\lambda_2$ fixed at 0.2. Since, $T^{zz}_{1N}$ = $-T^{xx}_{1N}~T^{yy}_{1N}$, therefore $T^{xx}_{1N}$ and $T^{yy}_{1N}$ are the only independent variables, with $|T^{zz}_{1N}| \leq |T^{xx}_{1N}|$ and $|T^{zz}_{1N}| \leq |T^{yy}_{1N}|$. Moreover, $T^{xx}_{1N}$ remains constant with the variation of $\lambda_1$.

Consider the region, $\lambda_1 \leq \lambda_2 = 0.2$, in Fig.~\ref{vary}. We see that $|T^{yy}_{1N}| \geq |T^{xx}_{1N}| \geq |T^{zz}_{1N}|$. 
Therefore, the $\cal{C}(\rho_{ij})$ is dependent only on $|T^{yy}_{1N}|$, and decreases with increasing $T^{yy}_{1N}$. Similarly, $\cal{I}(\rho_{ij})$ is varies with $|T^{yy}_{1N}|$ ($|T^{xx}_{1N}|$ is constant) and decreases with an identical rate, thus allowing $\mathcal{D}_L$ to remain frozen for $\lambda_1 \leq \lambda_2 = 0.2$, as observed in the inset of Fig.~\ref{vary}. For $\lambda_1 > \lambda_2 = 0.2$, $|T^{xx}_{1N}| \geq |T^{yy}_{1N}| \geq |T^{zz}_{1N}|$, and $\cal{C}(\rho_{ij})$ is dependent only on $|T^{xx}_{1N}|$, which is constant. Therefore, $\cal{C}(\rho_{ij})$ is constant for $\lambda_1 > \lambda_2 = 0.2$, but the decreasing $\cal{I}(\rho_{ij})$ forces $\mathcal{D}_L$ to decrease, leading to breakdown of freezing. The behavior is consistent with that observed in the phenomena of temporal freezing \cite{Smanis}.

For entanglement, no adiabatic freezing occurs. Since 0 $< |T^{\alpha\alpha}_{1N}| <$ 1, $\forall~\alpha$, the long-range concurrence is given by the relation, $\cal{E}_L$ = $\max\left[0, 1/2(|g^{+}_{1N}| - h^{+}_{1N})\right]$, where, $|g^{+}_{1N}|$ = $|T^{xx}_{1N}+T^{yy}_{1N}|$, and $h^{+}_{1N}$ = $1~ -~ |T^{xx}_{1N}||~T^{yy}_{1N}| >$ 0. As $\lambda_1$ increases $|g^{+}_{1N}|$ and $h^{+}_{1N}$ decreases, due to the fact that $|T^{yy}_{1N}|$ decreases, and $T^{xx}_{1N}|$ is constant. For, $|g^{+}_{1N}| > h^{+}_{1N}$,  $\cal{E}_L$ = $1/2(|g^{+}_{1N}| - h^{+}_{1N})$ and decreases with $\lambda_1$. For $|g^{+}_{1N}| \leq h^{+}_{1N}$, $\cal{E}_L$ = 0, and long-range entanglement vanishes. The above analysis can be compared with known results on quasi long-range entanglement. For $\lambda_1$ = $\lambda_2$, the system is isotropic, and we have $T^{xx}_{1N}$ = $T^{yy}_{1N}$ = $z$ (say). Moreover, $T^{zz}_{1N}$ = $-z^2$, which gives us the relation for the long-range entanglement, $\cal{E}_L$ = $\max\left[0, 1/2(z^2+2|z|-1))\right]$. Now for $x$ = $-\langle S_1^x S_N^x + S_1^y S_N^y\rangle$ = $-1/2~ z$, the above expression for entanglement reduces to $\cal{E}_L$ = $2\max\left[0, (x^2+|x|-1/4))\right]$, as shown in \cite{Sven}.

\section{Freezing of energy gap and quantum discord, at large $N$}

An important result discussed in the main text of the letter, is the adiabatic freezing of the energy gap, and its complementary relation to the freezing of quantum discord. In particular, it is shown that for the Hamiltonian in Eq.~(\ref{Aeq:ham3}), the quasi long-range quantum discord between the end spins in the ground state of the system is frozen for $\lambda_1 \leq \lambda_2$, while the energy gap remains constant in the complementary region $\lambda_1 \geq \lambda_2$.

Let us briefly describe the presence of energy gap in the spin Hamiltonian considered in the study. As mentioned in the main text, for the case, $\lambda_1$ = $\lambda_2$ =$\lambda$, the dispersion relation of the excitation energy is given by $\Delta_k$ = $\cos(k)$, where $k$ is the quasimomentum modes. These modes satisfy the following eigenvalue equation
$
\cot(k)[\cot((N-1)k/2)] = \lambda^2/(2-\lambda^2)
$, for $\lambda \neq$ 1, where the positive eigenstate parity has been considered \cite{Sven}.
The energy gap is then given by $k'$, which minimizes the dispersion relation $\Delta_k$ = $\cos(k')$, while satisfying the above eigenvalue equation.

\begin{figure}[t!]
%\vspace{0.5cm}
\epsfig{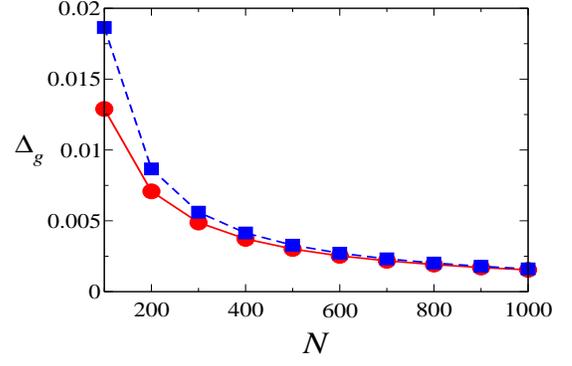}
\caption{(Color online.) 
%Understanding adiabatic freezing. 
Variation of the analytically and numerically estimated values of the energy gap with increasing size of the spin chain. 
The value of energy gap, using the analytical expression, is given by $\Delta_g = \min[f(\lambda_1), f(\lambda_2)]$ (blue-square), where $f(\lambda)$ is defined in Eq.~(\ref{Adelta}), and for exact numerical calculations (red-circle), for $\lambda_1$ = 0.4 and $\lambda_2$ = 0.6 (and, $\lambda_1$ = 0.6 and $\lambda_2$ = 0.4), for large $N$. The figure shows that at large $N$, $\Delta_g$ scales linearly as $1/N$.
}
\label{gap-scale}
%\end{center}
\end{figure}

\begin{figure}[h]
%\begin{center}
%\vspace{0.5cm}
\epsfig{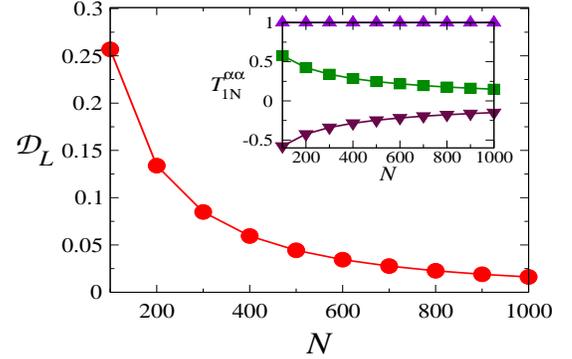}
 \caption{(Color online.) 
%Understanding adiabatic freezing. 
Variation of the long-range quantum discord (red-circle) with increasing size of the spin chain. The inset shows the behavior of the correlation functions, $T^{xx}_{1N}$ (green-square), $T^{zz}_{1N}$ (maroon-down-triangle), and $T^{yy}_{1N}$ (violet-up-triangle). $\lambda_1$ and $\lambda_2$ are set at 0.01 and 0.1, respectively.
The figure shows that at large $N$, $\cal{D}_L$ scales as $1/N$.
}
\label{discord-scale}
%\end{center}
\end{figure}

In large $N$ limit, for $k' = \pi/2-\delta$, where $\delta\rightarrow 0$, the dispersion relation provides an analytical expression for the energy gap ($\Delta_g$) as
\begin{eqnarray}
\Delta_g \approx \frac{\pi}{2N}\left(1 + \frac{2}{N(\lambda^2/(2-\lambda^2)+2}\right) = f(\lambda)
\label{Adelta}
\end{eqnarray}

%Using {mathematical} simulations, one can show that t
Numerical analysis for the case, $\lambda_1 \neq \lambda_2$, shows that the quasimomenta $k'$ corresponding to the energy gap satisfies the eigenvalue equation for $\lambda$ = $\min[\lambda_1,\lambda_2]$.
Thus, the energy gap is given by the analytical relation, $\Delta_g$ = $\min[f(\lambda_1), f(\lambda_2)]$, where $f(\lambda)$ is defined in Eq.~(\ref{Adelta}). 
Figure~\ref{gap-scale}, shows the agreement between the analytical expression for $\Delta_g$ and exact numerical calculations for large $N$. 
One can then show that $f(\lambda_1)<f(\lambda_2)$, for $\lambda_1<\lambda_2$, and $f(\lambda_2)\leq f(\lambda_1)$, for $\lambda_2\leq \lambda_1$. The adiabatic freezing of the energy gap, as $\lambda_1$ is varied, is thus evident for $\lambda_2 \leq  \lambda_1$, as $\Delta_g$ is independent of $\lambda_1$ in this range. 
%Moreover, numerical analysis show that even for moderate $N$, one can define
%$\Delta_g = \min[\Delta_g(\lambda_1), \Delta_g(\lambda_2)]$, where $\Delta_g(\lambda)$ is the numerically estimated energy gap for the balanced case, with coupling $\lambda$. The freezing of the energy gap is thus observed for all $N$, as shown in the main text of the letter.

It is known that for the model considered in the study, given by Eq.~(\ref{Aeq:ham3}), the long-range quantum correlation between the end spins is quasi long-range \cite{Sven}, i.e., the long-range quantum correlation vanishes with increasing $N$. The behavior of $\cal{D}_L$ with increasing system size is shown in Fig.~\ref{discord-scale}, which plots the value of the frozen quantum discord with increasing system size. The figure shows that in the large $N$ limit, both the frozen energy gap and quasi long-range quantum discord scales with $1/N$.


\begin{thebibliography}{99}
\bibitem{ent} {{Horodecki R.}, %\etal,
{Horodecki  P., Horodecki M.} and {Horodecki K.}, 
\textit{Rev. Mod. Phys.}, {\bf 81} (2009) 865.}

\bibitem{qcor} {{Modi K.}, %\etal, 
{Brodutch A., Cable H., Paterek T.} and {Vedral V.}, 
\textit{Rev. Mod. Phys.}, {\bf 84} (2012) 1655.}

\bibitem{disc} {{Henderson L.} and {Vedral V.}, \textit{J. Phys. A}, {\bf 34} (2001) 6899; 
{Ollivier H. and Zurek W. H. }, \textit{Phys. Rev. Lett.}, {\bf 88} (2002) 017901.}

\bibitem{comp} {{Datta A., Shaji A.} and { Caves C. M.}, \textit{Phys. Rev. Lett.}, {\bf 100} (2008) 050502; {Lanyon B. P.}, %\etal, 
{Barbieri M., Almeida M. P.}, and {White A. G.}, 
\textit{Phys. Rev. Lett.}, {\bf 101}, (2008) 200501.}

\bibitem{comp1}{{Madhok V.} and {Datta A.}, \textit{Phys. Rev. A}, {\bf 83} (2011) 032323; 
{Cavalcanti D.}, %\etal, 
{Aolita L., Boixo S., Modi K., Piani M.} and { Winter A.}, 
\textit{Phys. Rev. A}, {\bf 83} (2011) 032324; {Pirandola S.}, \textit{Sci. Rep.}, \textbf{4} (2014) 6956.}

\bibitem{comp2}{{Daki{\'c} B.} \etal, 
%Lipp Y. O., Ma X., Ringbauer M., Kropatschek S., Barz S., Paterek T., Vedral V., Zeilinger A., Brukner {\u C}.} and {Walther P.}, 
\textit{Nat. Phys.}, {\bf 8} (2012) 666; 
{Gu M.} \etal, 
%Chrzanowski H. M., Assad S. M., Symul T., Modi K., Ralph T. C.,  Vedral V.} and {Lam P. K.}, 
\textit{Nat. Phys.}, {\bf 8} (2012) 671.}

\bibitem{qpt} {{Dillenschneider R.}, \textit{Phys. Rev. B}, {\bf 78} (2008) 224413; 
{Sarandy M. S.}, \textit{Phys. Rev. A}, {\bf 80} (2009) 022108; 
%{Chen Y.-X.} and {Li S.-W.}, Phys. Rev. A, {\bf 81}  (2010) 032120; 
{Werlang T.}, %\etal,
{Ribeiro G. A. P.} and {Rigolin G.}, 
\textit{Phys. Rev. Lett.}, {\bf 83} (2011) 062334; 
%{Pal A. K.} and {Bose I.}, \textit{J. Phys. B: At. Mol. Opt. Phys}., {\bf 44}  (2011) 045101;
{Liu B. -Q.}, %\etal,
{Shao B., Li J.-G., Zou J.} and {Wu L.-A.}, 
\textit{Phys. Rev. A}, {\bf 83} (2011) 052112; 
{Allegra M.}, %\etal,
{Giorda P.} and {Montorsi A.}, 
\textit{Phys. Rev. B}, {\bf 84} (2011) 245133.}

\bibitem{mb} {{Werlang T.} and {Rigolin G.}, \textit{Phys. Rev. A}, {\bf 81}  (2011) 044101; 
{Campbell S.} \etal,
%, Apollaro T. J. G., Franco C. D., Banchi L., Cuccoli  A., Vaia R.,  Plastina F.} and {Paternostro M.}, 
\textit{Phys. Rev. A}, {\bf 84} (2011) 052316; 
{Mazzola L.} and {Paternostro M.}, \textit{Sci. Rep.}, {\bf 1} (2011) 199; 
{Tomasello B.}, %\etal,
Rossini D., Hamma A. and Amico L.,
\textit{EPL}, \textbf{96} (2011) 27002;
{Amico L.}, Rossini D., Hamma A. and Korepin V. E.,
\textit{Phys. Rev. Lett.}, \textbf{108} (2012) 240503};
{Prabhu R.}, %\etal,
{Sen(De) A.} and {Sen U.}, 
\textit{Phys. Rev. A}, {\bf 86} (2012) 012336; 
{Maziero J.}, %\etal,
{C{\'e}leri L. C., Serra R. M.} and {Sarandy M. S.}, 
\textit{Phys. Lett. A}, {\bf 376} (2012) 1540; 
{Pal A. K.} and {Bose I.}, \textit{Eur. Phys. J. B}, {\bf 85} (2012) 36; 
\textit{Eur. Phys. J. B}, {\bf 85} (2012) 277; 
{Mishra U.}, %\etal,
{Prabhu R., Sen(De) A.} and {Sen U.}, 
\textit{Phys. Rev. A}, {\bf 87} (2013) 052318; 
{{Ciliberti L.}, %\etal,%, 
Canosa N., and Rossignoli R.
\textit{Phys. Rev. A}, \textbf{88} (2013) 012119; 
{Cianciaruso M.} \etal, 
%Giampaolo S. M., Roga W., Zonzo G., Blasone M., and Illuminati F., 
arXiv:1412.1054};
{Huang Y.}, \textit{Phys. Rev. B}, {\bf 89}  (2014) 054410. 


\bibitem{epl} {{Dhar H. S.}, %\etal, 
{Ghosh R., Sen(De) A.} and {Sen U}, 
\textit{EPL}, {\bf 98} (2012) 30013; 
%{Dhar H. S., Ghosh R., Sen(De) A.} and {Sen U.}, 
\textit{Phys. Lett. A}, {\bf 378} (2014) 1258. }

\bibitem{qbio} {{Br{\'a}dler K.}, %\etal,
{Wilde M. M., Vinjanampathy S.} and {Uskov D. B.}, 
\textit{Phys. Rev. A}, {\bf 82} (2010) 062310;
{Chanda T.}, % \etal,
{Mishra U., Sen(De) A.} and {Sen U.}, 
arXiv:1412.6519.}



\bibitem{qmet} {{Modi K.}, %\etal,
{Cable H., Williamson M.} and {Vedral V.}, 
\textit{Phys. Rev. X}, {\bf 1} (2011) 021022.}

\bibitem{qmet1} {{Girolami D.}, %\etal,
{Tufarelli T.} and {Adesso G.}, 
\textit{Phys. Rev. Lett.}, {\bf 110} (2013) 240402; 
{Girolami D.} \etal,
%, Souza A. M., Giovannetti V., Tufarelli T., Filgueiras  J. G., Sarthour R. S.,  Soares-Pinto D. O., Oliveira I. S.} and {Adesso G.}, 
\textit{ibid.}, {\bf 112} (2014) 210401.}


\bibitem{swap}{{{\.Z}ukowski M.}, %\etal,
{Zeilinger A., Horne M. A.} and {Ekert A. K.}, 
\textit{Phys. Rev. Lett.}, {\bf 71} (1993) 4287; 
{\.Zukowski M., Zeilinger A.} and {Weinfurter H.}, 
\textit{Annals N.Y. Acad. Sci.}, \textbf{755} (1995) 91; 
{Bose S.}, %\etal,
{Vedral V.} and {Knight P. L.}, 
\textit{Phys. Rev. A}, \textbf{57} (1998) 822.}
%{Bose S., Vedral V.} and {Knight P. L.}, 
%\textit{Phys. Rev. A}, \textbf{60} (1999) 194.}

\bibitem{repeat}{{Briegel H.-J.}, %\etal,
{D{\"u}r W., Cirac J. I.} and {Zoller P.}, 
\textit{Phys. Rev. Lett.}, {\bf 81}  (1998) 5932.}

%\bibitem{concentrate}C. H. Bennett, H. J. Bernstein, S. Popescu, and B. Schumacher, Phys. Rev. A {\bf 53}, 2046 (1996).

\bibitem{local}{{Verstraete F.}, %\etal,
{Popp M.}, and {Cirac J. I.}, 
\textit{Phys. Rev. Lett.}, {\bf 92} (2004) 027901;
{Popp M.}, %\etal,
{Verstraete F., Mart{\'i}n-Delgado  M. A.} and { Cirac J. I.}, 
\textit{Phys. Rev. A}, {\bf 71} (2005) 042306.} 
%%{Popescu S.} and {Rohrlich D.}, \textit{Phys. Lett. A}, \textbf{166} (1992) 293; 
%%{Sen(De) A.} \etal,
%%%, Sen U., Wiesniak M., Kaszlikowski D.} and { \.Zukowski M.}, 
%%\textit{Phys. Rev. A}, \textbf{68} (2003) 062306.}
%%
%%
\bibitem{grudka}{{ W{\'o}jcik A.}, %\etal,
{{\L}uczak T., Kurzy{\'n}ski P., Grudka A., Gdala T.} and {Bednarska M.}, 
\textit{Phys. Rev. A}, {\bf 72} (2005) 034303.}

\bibitem{venuti}{{Venuti L. C.}, %\etal},
{Degli E. B. C.} and {Roncaglia M.}, 
\textit{Phys. Rev. Lett.}, {\bf 96} (2006) 247206;
%{Campos V. L., Degli E. B. C.} and { Roncaglia M.}, 
\textit{ibid.}, {\bf 99} (2007) 060401;
%
%\bibitem{zan} 
{Venuti L. C.}, %\etal,
{Giampaolo S. M.,  Illuminati F.} and {Zanardi P.}, 
\textit{Phys. Rev. A}, {\bf 76} (2007) 052328.}



\bibitem{ilu1} {{Giampaolo S. M.} and { Illuminati F.}, \textit{Phys. Rev. A}, {\bf 80} 050301 (2009); \textit{New J. Phys.}, {\bf 12} (2010) 025019.}

\bibitem{bose} {{Yang S.}, %\etal,
{Bayat A.} and {Bose S.}, 
\textit{Phys. Rev. A}, 84 (2011) 020302(R);
{Bayat A.}, %\etal,
{Sodano P.} and {Bose S.}, 
\textit{Quant. Inf. Proc.}, {\bf 11} (2012) 89.}
%{Rafiee M.} and {Mokhtari H.}, \textit{Phys. Rev. A}, {\bf 87} (2013) 022304.}

{\bibitem{referee}{{Son W.}, Amico L., Plastina F., and Vedral V.,
\textit{Phys. Rev. A}, \textbf{79} (2009) 022302.}}

\bibitem{open1} {{Breuer H.-P.} and { Petruccione F.}, \emph{The Theory of Open Quantum Systems} (Oxford University Press, Oxford) 2002.}

\bibitem{robust} {{Werlang T.}, %\etal,
{Souza S., Fanchini F. F.} and {Villas Boas C. J.}, 
\textit{Phys. Rev. A}, {\bf 80} (2009) 024103;
{Lo Franco R.}, %\etal,
{Bellomo B., Maniscalco S.} and {Compagno G. }, 
\textit{Int. J. Mod. Phys. B}, {\bf 27} (2013) 1345053, and references therein.}

\bibitem{esd} {{Yu T.} and {Eberly J. H.},
\textit{Phys. Rev. Lett.}, \textbf{93} (2004) 140404;
{Almeida M. P.}, %\etal, 
{de Melo F., Hor-Meyll M., Salles A., Walborn S. P.,  Ribeiro P. H. S.} and {Davidovich L.}, 
\textit{Science}, {\bf 316}  (2007) 579.}
%{Yu T.} and {Eberly J. H.}, Science, {\bf 323} (2009) 598.}


\bibitem{manis} {Mazzola L.}, %\etal,
{Piilo J.} and { Maniscalco S.}, 
\textit{Phys. Rev. Lett.}, {\bf 104} (2010) 200401.

\bibitem{expt} {Xu J.-S.}, %\etal, 
{Xu X.-Y., Li C.-F., Zhang C.-J., Zou X.-B.} and {Guo G.-C.}, 
\textit{Nat. Comm.}, \textbf{1} (2010) 7;
%%{Cornelio M. F.} {\etal}, \textit{Phys. Rev. Lett.}, \textbf{109} (2012) 190402;
%%{Xu J.-S.} \etal, 
%Sun K., Li C.-F., Xu X.-Y., Guo G.-C.,	Andersson E., Lo Franco R.}, and	{Compagno G.}, 
%\textit{Nat. Comm.} \textbf{4} (2013) 2851;
%{Paula F. M.} {\etal}, \textit{Phys. Rev. Lett.}, \textbf{111} (2013) 250401.}
{Auccaise R.} \etal,
% C{\'e}leri  L. C., Soares-Pinto D. O. , deAzevedo E. R., Maziero  J., Souza A. M., Bonagamba T. J., Sarthour R. S., Oliveira I. S.} and {Serra R. M.}, 
\textit{Phys. Rev. Lett.}, {\bf 107} (2011) 140403;
{Silva I. A.} \etal, 
%Girolami D., Auccaise R., Sarthour R. S., Oliveira I. S., Bonagamba T. J., deAzevedo E. R., Soares-Pinto D. O.} and {Adesso G.}, 
\textit{Phys. Rev. Lett.}, \textbf{110}  (2013) 140501.

\bibitem{freezing} {{Haikka P.}, %\etal,
{Johnson T. H.} and {Maniscalco S.}, 
\textit{Phys. Rev. A}, {\bf 87}  (2013) 010103(R);
{Mazzola L.}, %\etal,
{Piilo J.} and {Maniscalco S.}, 
\textit{Int. J. Quant. Inf.}, {\bf 09}  (2011) 981;
%R. Lo Franco, B. Bellomo, E. Andersson, and G. Compagno, Phys. Rev. A {\bf 85}, (2012) 032318;
%B. You and L.-X. Cen, Phys. Rev. A {\bf 86}, (2012) 012102;
{Bellomo B.}, %\etal,
{Lo Franco R.} and {Compagno G.}, 
\textit{Phys. Rev. A}, \textbf{86}  (2012) 012312;
%B. Aaronson, R. L. Franco, and G. Adesso, Phys. Rev. A {\bf 88}, (2013) 012120;
{Xu J.-S.} \etal,
%, Sun K., Li C.-F., Xu X.-Y., Guo  G.-C., Andersson  E., Lo Franco R.} and {Compagno G.}, 
\textit{Nat. Commun.}, \textbf{4}, (2013) 2851;
%B. Aaronson, R. Lo Franco, G. Compagno, and G. Adesso, New J. Phys. \textbf{15},  (2013) 093022;
{Montealegre J. D.}, %\etal, 
{Paula F.M., Saguia A.} and {Sarandy M. S.}, 
\textit{Phys. Rev. A}, {\bf 87}  (2013) 042115;
%
%T. R. Bromley, M. Cianciaruso, R. Lo Franco, and G. Adesso, J. Phys. A: Math. Theor. \textbf{47},  (2014) 405302;
{Cianciaruso M.}, %\etal,
{Bromley T. R., Roga W., Lo Franco R.} and {Adesso G.}, 
%\emph{Universality of the freezing of geometric quantum correlations}, 
\textit{Sci. Rep.}, \textbf{5}, (2015) 10177;
{Bromley T. R.}, %\etal,
{Cianciaruso M.} and {Adesso G.}, 
\textit{Phys. Rev. Lett.}, {\bf 114}  (2015) 210401.}



\bibitem{titas}{{Chanda T.}, %\etal,
{Pal A. K., Biswas A., Sen(De) A.}, and {Sen U.}, 
%\emph{To freeze or not to: Quantum correlations under local decoherence}, 
\textit{Phys. Rev. A}, {\bf 91}  (2015) 062119.}

%%\bibitem{ger-coh} {{Bromley T. R.} \etal,
%%% Cianciaruso M.} and {Adesso G.}, 
%%\textit{Phys. Rev. Lett.}, {\bf 114}  (2015) 210401. }

\bibitem{freeze-ent}{{Carnio E. G.}, %\etal, 
{Buchleitner A.} and {Gessner M.}, 
\textit{Phys. Rev. Lett.}, \textbf{115}  (2015) 010404.}

\bibitem{new} {{Sahling S.} \etal, 
%Remenyi G., Paulsen C., Monceau P. , Saligrama V., Marin C., Revcolevschi A., Regnault L. P., Raymond S.} and { Lorenzo J. E.}, 
\textit{Nat. Phys.}, {\bf 11} (2015), 255.}

\bibitem{iontrap} {{Zippilli S.}, %\etal, 
{Johanning M., Giampaolo S. M.,  Wunderlich Ch.} and {Illuminati F.}, 
\textit{Phys. Rev. A}, {\bf 89} (2014) 042308.}

\bibitem{flux}{{Zippilli S.},  %\etal,
{Grajcar M., Il{'}ichev E.} and {Illuminati F.}, 
\textit{Phys. Rev. A}, \textbf{91}  (2015) 022315.}


\bibitem{expt3} Aenesen M. C. %\etal, 
{Bose S.} and {Vedral V.}, 
\textit{Phys. Rev. Lett.}, {\bf 87} (2001) 017901;
%\bibitem{expt1} 
{Wiesniak M.}, %\etal, 
{Vedral V.} and {Brukner C.}, 
\textit{New J. Phys.}, {\bf 7} (2005) 258;
%{Brukner C., Vedral V.} and {Zeilinger A.}, Phys. Rev. A, {\bf 73} (2006) 012110; 
%{Wiesniak M., Vedral V.} and {Brukner C.}, Phys. Rev. B, {\bf 78} (2008) 064108.}
%\bibitem{expt2} {
{Lima Sharma A. L.} and {Gomes A. M.}, \textit{EPL}, \textbf{84} (2008) 60003; 
{Chakraborty T.}, %\etal, 
{Singh H., Singh S., Gopal R. K.}, and {Mitra C.}, 
\textit{J. Phys. Condens. Matter}, {\bf 25} (2013) 425601; 
{Chakraborty T.}, %\etal, 
{Sen T. K., Singh H., Das D., Mandal S. K.} and {Mitra C.}, 
\textit{J. Appl. Phys.}, {\bf 114} (2013) 144904; 
%%{Das D.} et al,
%%%, Singh H., Chakraborty T., Gopal R. K.} and { Mitra C.}, 
%%\textit{New J. Phys.}, {\bf 15} (2013) 013047; 
%%{Singh H.} et al, 
%%%Chakraborty T., Das D.,  Jeevan H. S., Tokiwa Y., Gegenwart P.} and {Mitra C.}, 
%%\textit{ibid.}, {\bf 15} (2013) 113001; 
{Chakraborty T.}, %\etal,
{Singh H.} and {Mitra C. }, 
\textit{ibid.}, {\bf 115} (2014) 034909.


\bibitem{LSM} {{Lieb E.}, % \etal, 
{Schultz T.} and {Mattis D.}, 
\textit{Ann. Phys.}, {\bf 16} (1961) 407.};
Barouch E., McCoy B. M. and Dresden M., 
\textit{Phys. Rev. A}, \textbf{2} (1970) 1075;
Barouch E. and McCoy B. M., \textit{Phys. Rev. A}, \textbf{3} (1971)
786.

%\bibitem{JW} {{Jordan P.} and {Wigner E.}, \textit{Z. Phys.}, {\bf 47} (1928) 631.}

\bibitem{supple} See Supplementary Material at the end of the manuscript for further details.

\bibitem{LUO} {{Luo S.}, \textit{Phys. Rev. A}, {\bf 77} 042303 (2008).}
%%%
%%%\bibitem{supple} See Supplemental Material for the methodology, mathematical
%%%%to solve the spin chain Hamiltonian, 
%%%description of adiabatic freezing, %in terms of two-site correlation functions
%%%and additional data and figures.





%%
%%
%%an illustrative description of the end and bulk spins in a chain, methodology used in solving the spin Hamiltonian and computing the quantum correlations, and extra set of plots to describe the variation of long-range discord, entanglement, and energy gap with the end couplings.

\bibitem{woot}{{Hill S.} and {Wootters W. K.}, \textit{Phys. Rev. Lett.}, {\bf 78} 5022 (1997); {Wootters W. K.}, \textit{ibid.}, {\bf 80} 2245 (1998).}\

\bibitem{sym-dis}{{Rulli C. C.} and {Sarandy M. S.}, \textit{Phys. Rev. A}, \textbf{84} 042109 (2011).}
%%; {Maziero J.} \etal,
%%%, Celeri L. C.} and {Serra R. M.}, 
%%arXiv:1004.2082.}\

\bibitem{anis-expt}{{{Cuccoli A.}}, %\etal, 
%in \textit{Experimental Magnetism}, eds. { Kalvius G. M.} and {Tebble R. S.}, %Vol. \textbf{1} 
%(Wiley, New York) 1979.}}
%Alessandro Cuccoli, 
Roscilde T., Vaia R., and Verrucchi P.,
\textit{Phys. Rev. Lett.}, \textbf{90} 167205 (2003).}
%%%{Soeya S.} \etal,
%%%%, Nakamura S., Imagawa T.} and {Narishige S.}, 
%%%\textit{J. Appl. Phys.}, \textbf{77} 5838 (1995);
%%%{Shimizu Y.} \etal, 
%%%%Horibe M., Nanba H., Takami T.} and {Itoh M.}, 
%%%\textit{Phys. Rev. B}, \textbf{82} 094430 (2010);
%%%{Sch{\"a}pers M.} {\etal}, \textit{ibid.}, \textbf{88} 184410 (2013), and references therein.}\


\bibitem{vencite} {{Chiara G. D.}, %\etal, 
{Brukner C., Fazio R., Palma G. M.} and {Vedral V.}, 
\textit{New J. Phys.}, {\bf 8} 95 (2006).}
%%\bibitem{LSM} E. Lieb, T. Schultz, and D. Mattis, Ann. Phys. {\bf 16}, 407 (1961).
%%
%%\bibitem{JW} P. Jordan and E. Wigner, Z. Phys. {\bf 47}, 631 (1928).
%%
%%
%%
%%\bibitem{LUO} S. Luo, Phys. Rev. A {\bf 77}, 042303 (2008).
%\bibitem{lieb} E. Lieb, T. Schultz, and D. Mattis, Ann. Phys. {\bf 16}, 407 (1961).

%\bibitem{egap-probe} {{Berkley A. J.} {\etal}, \textit{Phys. Rev. B}, \textbf{87} (2013) 020502(R).}



\bibitem{QWD} {{Horodecki M.} \etal, 
%{Horodecki P., Horodecki R.,  Oppenheim J., Sen(De) A., Sen U.} and {Synak-Radtke B.},
\textit{Phys. Rev. A}, \textbf{71} (2005) 062307.}

%%{\bibitem{MG}  {{Majumdar C. K.} and  {Ghosh D. K.}, \textit{J. Math. Phys.}, \textbf{10} (1969) 1388.}
%%
%%\bibitem{AKLT} {{Affleck I.} \etal,
%%%   T.   Kennedy,   E. H.   Lieb,   and   H.   Tasaki,
%%\textit{Commun. Math. Phys.}, \textbf{115} (1988) 477.}
%%
%%\bibitem{gapped} {{Affleck I.} \etal,
%%\textit{Phys.  Rev.  Lett.}, \textbf{59} (1987) 799;
%%{Chen X.} \etal,
%%%, Zheng-Cheng Gu, Xiao-Gang Wen, Classification of Gapped Symmetric Phases in 1D Spin Systems 
%%\textit{Phys. Rev. B}, \textbf{83} (2011) 035107.}}$J_1






% %  
\end{thebibliography}

\begin{thebibliography}{99}

\bibitem{SLSM} {{Lieb E.}, %\emph{et al.},
{Schultz T.} and {Mattis D.}, 
\textit{Ann. Phys.}, {\bf 16} (1961) 407.}

\bibitem{SJW} P. Jordan and E. Wigner, Z. Phys. {\bf 47}, 631 (1928).

\bibitem{Sdisc} {{Henderson L.} and {Vedral V.}, \textit{J. Phys. A}, {\bf 34} (2001) 6899; 
{Ollivier H. and Zurek W. H. }, Phys. Rev. Lett., {\bf 88} (2002) 017901.}



%\bibitem{JW} {{Jordan P.} and {Wigner E.}, \textit{Z. Phys.}, {\bf 47} (1928) 631.}

\bibitem{SLUO} {{Luo S.}, \textit{Phys. Rev. A}, {\bf 77} 042303 (2008).}
%%%

\bibitem{Swoot}{{Hill S.} and {Wootters W. K.}, \textit{Phys. Rev. Lett.}, {\bf 78} (1997)  5022; {Wootters W. K.}, \textit{Phys. Rev. Lett.}, {\bf 80} (1998) 2245.}

%\bibitem{QWD} M. Horodecki, P. Horodecki, R. Horodecki, J. Oppenheim, A. Sen(De), U. Sen, and B. Synak-Radtke, Phys. Rev. A \textbf{71}, 062307 (2005).

\bibitem{Smanis} {{Mazzola L.}, %\emph{et al.},
{Piilo J.} and { Maniscalco S.}, 
\textit{Phys. Rev. Lett.}, {\bf 104} (2010) 200401.}

\bibitem{Sven}{{Venuti. L. C.} %\emph{et al.},
{Giampaolo S. M.,  Illuminati F.} and {Zanardi P.}, 
\textit{Phys. Rev. A}, {\bf 76} (2007) 052328.}
% %  
\end{thebibliography}
\end{document}